\documentclass[aps,prc,twocolumn,groupedaddress]{revtex4-1}

\usepackage{amssymb}

\usepackage{color}
\usepackage{amsmath}
\usepackage{multirow}
\usepackage{booktabs}
\usepackage{bm}
\usepackage{braket}
\usepackage{amsmath}
\usepackage{graphicx}
\usepackage{ulem}

\usepackage[pdfstartview=FitH, colorlinks,
 linkcolor={red!60!black!},
 anchorcolor={green!80!black!},
 urlcolor={blue!80!black!},
 citecolor={blue!50!black!}]{hyperref}
\usepackage[table]{xcolor}

\begin{document}





\title{Unveiling potential neutron halos in intermediate-mass nuclei: an \textit{ab initio} study}


\author{H. H. Li}
\author{J. G.  Li}\email[]{jianguo\_li@impcas.ac.cn}
\author{M. R. Xie}
\author{W. Zuo}

\affiliation{CAS Key Laboratory of High Precision Nuclear Spectroscopy, Institute of Modern Physics,
Chinese Academy of Sciences, Lanzhou 730000, China}
\affiliation{School of Nuclear Science and Technology, University of Chinese Academy of Sciences, Beijing 100049, China}

\date{\today}

\begin{abstract}

Halos epitomize the fascinating interplay between weak binding, shell evolution, and deformation effects, especially in nuclei near the drip line. In this Letter, we apply the state-of-the-art \textit{ab initio} valence-space in-medium similarity renormalization group approach to predict potential candidates for one- and two-neutron halo in the intermediate-mass region. Notably, we use spectroscopic factors (SF) and two-nucleon amplitudes (TNA) as criteria for suggesting one- and two-neutron halo candidates, respectively. This approach is not only theoretically sound but also amenable to experimental validation. Our research focuses on Mg, Al, Si, P, and S neutron-drip-line nuclei, offering systematic predictions of neutron halo candidates in terms of separation energies, SF (TNA), and average occupation. The calculation suggests the ground states of $^{40,42,44,46}$Al, $^{41,43,45,47}$Si, $^{46,48}$P, and $^{47,49}$S are promising candidates for one-neutron halos, while $^{40,42,44,46}$Mg, $^{45,47}$Al, $^{46,48}$Si, $^{49}$P, and $^{50}$S may harbor two-neutron halos. In addition, the relative mean-square neutron radius between halo nuclei and \textit{inner core} is calculated for suggested potential neutron halos. Finally, the relations of halo formations and shell evolution are discussed.

\end{abstract}

\pacs{}



\maketitle

\textit{Introduction.}~
Nuclear halo formation is an important feature of nuclei with extreme $N/Z$ ratios near the limits of nuclear stability. 
Only a few neutron halo nuclei have been identified at the light nuclear drip line,
such as $^6$He~\cite{PhysRevLett.93.142501, PhysRevLett.108.052504}, $^{11}$Li~\cite{PhysRevLett.55.2676}, $^{14}$Be~\cite{PhysRevC.64.061301}, $^{17}$B~\cite{PhysRevLett.31.614}, $^{22}$C~\cite{PhysRevLett.104.062701}, and $^{29}$F~\cite{PhysRevLett.124.222504} two-neutron ($2n$) halos, alongside the one-neutron ($1n$) halo nuclei including $^{11}$Be~\cite{PhysRevLett.74.30}, $^{15,19}$C~\cite{PhysRevC.69.034613, PhysRevLett.106.172701, PhysRevLett.74.3569, PhysRevLett.83.1112}, $^{31}$Ne~\cite{PhysRevLett.112.142501}, and $^{37}$Mg~\cite{PhysRevLett.112.242501}. The discovery and ongoing research into neutron halos have significantly altered the foundational concepts of nuclear physics, highlighting large sizes, soft $E1$ transitions, narrow parallel momentum distributions of the residue, large cross sections for nucleon(s) breakup, core shadowing, sudden and large increase of the reaction cross section~\cite{TANIHATA2013215}.
Halo nuclei are characterized by their unique structure, featuring one or two nucleons, typically neutrons, that are weakly bound and decoupled from the nuclear core~\cite{PhysRevLett.112.142501}, which can be treated in a core plus one or two nucleons picture.
Their significance in nuclear physics stems from their ability to shed light on new facets of nucleon-nucleon interactions, challenge traditional nuclear models, and expand our understanding of nuclei under extreme conditions.

Halo nuclei are predominantly influenced by valence $1n$ or $2n$  that are weakly bound. 
Another necessary condition for halo formation is that the valence $1n$ or $2n$ occupy a low orbital angular momentum with $ \ell =0$ or 1. This low-$\ell$ facilitates the extensive spread of the neutron wave function far from the core, owing to the minimal, or absent, centrifugal barrier~\cite{TANIHATA2013215}.
This phenomenon, shown by nuclei such as the heaviest neutron halo nucleus $^{37}$Mg, exemplifies a deformed $p$-wave halo with $S_n = 0.22^{+0.12}_{-0.09}$ MeV~\cite{PhysRevLett.112.242501, PhysRevC.90.061305}, highlighting its halo characteristics.


The emergence of halo structures signals profound alterations in the shell configuration, wherein the mixing of intruder states can precipitate substantial deformations~\cite{PhysRevLett.103.262501, PhysRevLett.112.142501}.
The characteristic halo of $^{11}$Li, for instance, can be traced back to the vanishing of the $N=8$ shell closure with the intruder $1s_{1/2}$ orbital that leads to strong mixing between $0p_{1/2}$ and $1s_{1/2}$ orbits~\cite{PhysRevLett.83.496, PhysRevLett.100.192502}.
The $2n$ halo structure observed in $^{29}$F emerges from the erosion of the traditional $N=20$ shell gap, due to the intrusion of the 1$p_{3/2}$ orbital from a higher shell.
The case of $^{31}$Ne was strongly suggestive of the presence of deformation~\cite{PhysRevLett.112.142501}, which is driven by a disappearance of the $N=20$ shell closure due to the near degeneracy of the $\nu 0f_{7/2}$ and $\nu 1p_{3/2}$ orbits ~\cite{PhysRevC.81.021304}. 
In addition, halo structures may impart added stability, affecting the location of the drip line, and deepening our understanding of nuclear structure~\cite{Tsunoda2020}.
The neutron-rich Mg, Al, Si, P, and S isotopes provide excellent conditions for the study of halo structures. 

Various theoretical models have been developed to describe and anticipate such features, including the few-body model~\cite{ZHUKOV1993151,doi:10.1146/annurev.ns.45.120195.003111}, shell model~\cite{PhysRevLett.70.1385, PhysRevLett.78.2708}, antisymmetrized molecular dynamics~\cite{PhysRevC.61.024303}, halo effective field theory~\cite{PhysRevC.89.014325}, density functional theory~\cite{TERASAKI1996371, PhysRevC.78.064305, PhysRevC.84.024311, PhysRevC.88.054305}, etc. The valence-space in-medium similarity renormalization group (VS-IMSRG)~\cite{PhysRevC.85.061304, HERGERT2016165, PhysRevLett.118.032502, PhysRevC.107.014302} stands out as a formidable \textit{ab initio} technique ~\cite{HERGERT2016165,doi:10.1146/annurev-nucl-101917-021120,WEGNER200177}. This method offers an efficient solution to tackle the intricacies of the nuclear many-body problem. Moreover, it boasts the advantage that no additional parameters are introduced in the calculation, ensuring reliable predictions.
In this study, we employ the VS-IMSRG to  suggest potential candidates for 
$1n$ and $2n$ halo nuclei within the intermediate-mass region. 
Diverging from traditional methodologies that rely on root-mean-square radii~\cite{LI2022137225, PhysRevC.85.024312} and density distribution~\cite{PhysRevC.101.031301, PhysRevC.82.011301, SUN20212072}, our strategy is to characterize the $1n$ and $2n$ halos utilizing spectroscopic factor (SF)~\cite{LI2022137225} and two-nucleon amplitudes (TNA)~\cite{PhysRevC.82.044616}, respectively.
This methodology offers a fresh perspective on suggesting and understanding halo nuclei.

This Letter is organized as follows. First, we introduce the theoretical framework of VS-IMSRG. We then proceed to calculate the single- ($S_n$) and two-neutron ($S_{2n}$) separation energies for neutron-rich Mg, Al, Si, P, and S isotopes, as well as other physical quantities such as SF and TNA. Finally, we present our suggestions for nuclei exhibiting characteristics of $1n$ and $2n$ halos and discuss the interplay between shell evolution and halo properties.

\textit{Method.}~
Beginning with an intrinsic $A$-nucleon Hamiltonian
\begin{equation}
H=\sum_{i<j}^{A}\left(\frac{(\boldsymbol{p}_{i} - \boldsymbol{p}_{j})^2}{2m A} + v_{i j}^{NN}\right)+\sum_{i<j<k}^{A} v_{i j k}^{3 N},
    \label{H_in}
\end{equation}
where $\boldsymbol{p}$ denotes the nucleon momentum in the laboratory, $m$ is the nucleon mass, $v^{NN}$ and $v^{3N}$ correspond to the nucleon-nucleon ($NN$) and three-nucleon ($3N$) interactions, respectively. The $\chi$EFT $NN + 3N$ interaction EM1.8/2.0~\cite{PhysRevC.83.031301, PhysRevC.93.011302} is employed in the present work. Additionally, calculations were also carried out using the $NN + 3N$ local and nonlocal (lnl) interaction~\cite{PhysRevC.101.014318}, for which the induced $3N$ force is neglected by adopting a large similarity renormalization group scale of $\lambda =$ 2.6 fm$^{-1}$ for the $NN$ interaction.  For both interactions, we take the harmonic-oscillator basis at $\hbar \omega =16$ MeV with $e_{\rm max}= 2n + l = 14$ and $E_{3\rm max} = 14$, ensuring sufficient convergence for our calculations \cite{YUAN2024138331}. Then, the Hamiltonian is rewritten as normal-ordered operators and typically truncated at the normal-ordered two-body level. In this way, the contribution of the $3N$ interaction can be naturally included in a normal-ordered two-body approximation~\cite{HERGERT2016165,PhysRevLett.109.052501}, while the residual normal-ordered three-body term is neglected.

The VS-IMSRG aims at decoupling the normal-ordered Hamiltonian from the large Hilbert space to a small valence space. This is achieved by solving the flow equation,
\begin{equation}
\frac{dH(s)}{ds}=[\eta(s),H(s)]
\label{FE}
\end{equation}
with an anti-Hermitian generator,
\begin{equation}
\eta(s)\equiv\frac{dU(s)}{ds}U^{\dagger}(s)=-\eta^{\dagger}(s).
\end{equation}

In the present work, the valence space of protons in the $sd$ shell and neutrons in the $pf$ shell above the $^{28}$O inner core is adopted. We use the Magnus formalism~\cite{PhysRevC.92.034331} of VS-IMSRG with ensemble normal ordering~\cite{PhysRevLett.118.032502}, in which the effects of $3N$ force can be captured at the two-body level, to generate consistently valence-space effective Hamiltonians. Subsequently, the obtained Hamiltonian can then be exactly diagonalized using the large-scale shell model code KSHELL~\cite{SHIMIZU2019372}, which is also used to calculate the SF and TNA.

\begin{figure*}[!htb]
\includegraphics[width=1.9\columnwidth]{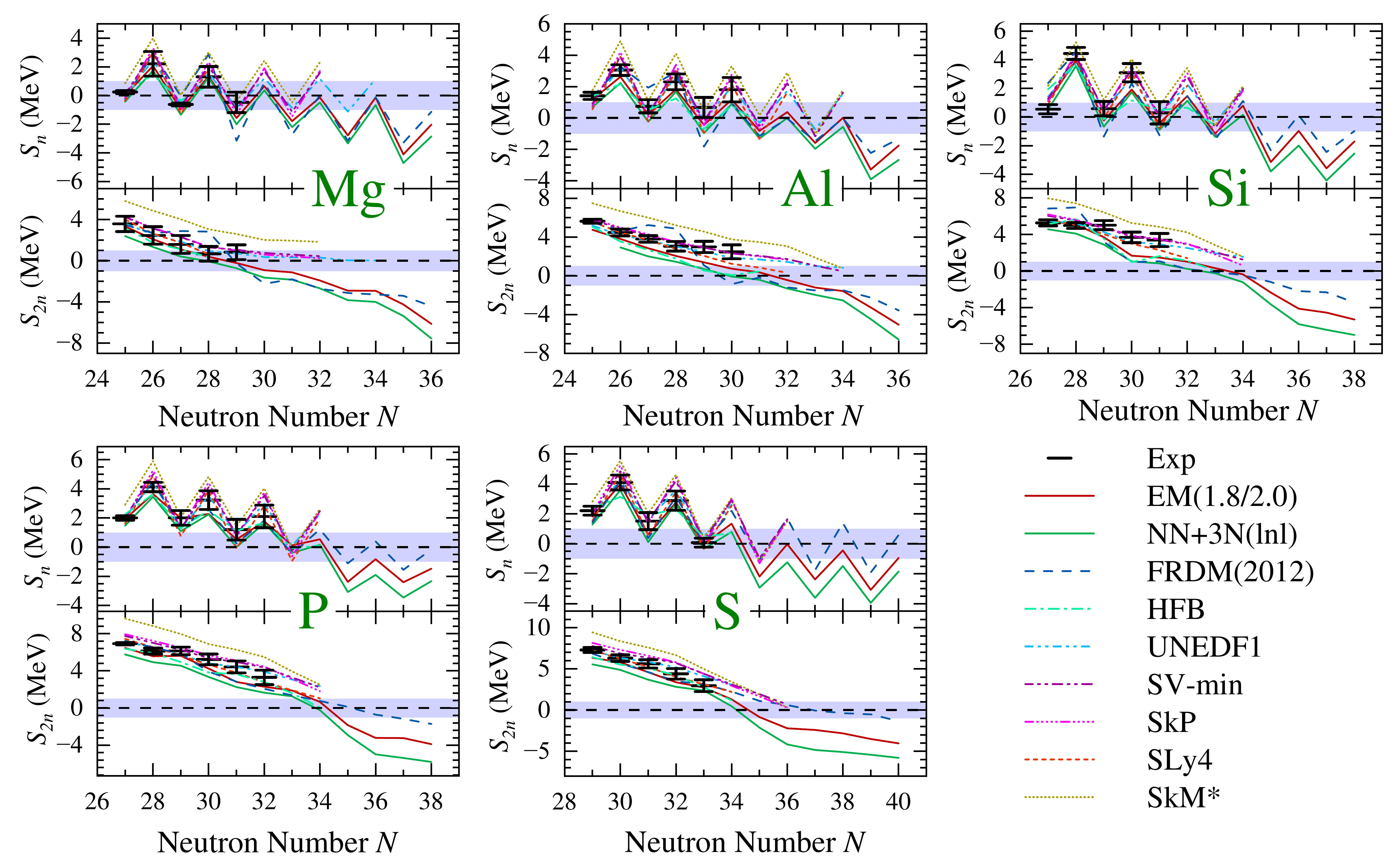}
\vspace{-0.3cm}
\caption{The calculated $S_n$ and $S_{2n}$ for Mg, Al, Si, P, and S isotopes using EM1.8/2.0 and $NN + 3N$(lnl) interactions, along with experimental data and results from other theoretical models. The highlighted regions indicate  $-1\leq S_n \leq 1$ and $-1\leq S_{2n} \leq 1.4$, which are key areas of interest.}
\label{Sn}
\end{figure*}

\textit{Results.}~
It is well recognized that weakly binding in low-$\ell$ components is a salient feature and a necessary condition in the formation of nuclear halos. Therefore, to foresee $1n$ and $2n$ halo nuclei, our first step entailed the calculation of the $S_n$ and $S_{2n}$ for $^{37-48}$Mg, $^{38-49}$Al, $^{41-52}$Si, $^{42-53}$P, and $^{45-56}$S. These calculations were performed utilizing the EM1.8/2.0 and $NN + 3N$(lnl) interactions within the VS-IMSRG framework. 
In our previous study~\cite{YUAN2024138331}, we applied the same nuclear forces to the $N = 28$ region and successfully reproduced crucial observables, such as low-lying spectra and deformations.
The results of calculated separation energies using \textit{ab initio} VS-IMSRG are summarized in Fig.~\ref{Sn}, along with available experimental data~\cite{Wang_2021,ensdf} and other theoretical results including the finite-range droplet model [FRDM(2012)]~\cite{MOLLER20161}, Gogny-Hartree-Fock-Bogoliubov (Gogny-HFB)~\cite{HFBgogny}, and Skyrme energy density functionals theory (DFT) with UNEDF1, SV-min, SkP, SLy4, and SkM$^{\ast}$~\cite{DFTSkyrme} potentials. 

In our study of potential $2n$ halo nuclei, we reference $^{17}$B and $^{29}$F as established examples with their $S_{2n}$ values recorded at 1.384(0.205) MeV~\cite{PhysRevLett.89.012501} and 1.126(0.539) MeV~\cite{PhysRevLett.109.202503}, respectively. To suggest potential halo candidates, we consider a broad range of low $S_{2n}$ values.
For potential $1n$ halos, we consider that $S_n$ should be smaller than 1 MeV. This threshold is slightly above the $S_n$ of about 900 keV for the $p$-wave halo in $^{29}$Ne~\cite{PhysRevC.93.014613,LI2022137225}.
Nuclei with $S_n$ and $S_{2n}$ larger than $-1$ MeV, classified as unbound, are still considered to be weakly bound and meet the condition for small separation energy necessary for neutron halos, considering theoretical uncertainties and the absence of continuum coupling.
Under this assumption, Fig.~\ref{Sn} highlights the regions with $-1\leq S_n \leq 1$ and $-1\leq S_{2n} \leq 1.4$~\cite{PhysRevLett.112.242501}. Nuclei in these shaded areas meet the criteria for being weakly bound neutron halos.

The odd-even staggering pattern of $S_n$ and the decreasing trend of $S_{2n}$ with increasing $N$ are well reproduced for those isotopes. 
For Mg isotopes, the $1n$ drip line is located at $^{37}$Mg, while $^{39}$Mg, confirmed experimentally, is unbound~\cite{PhysRevC.94.054302, Baumann2007}.
Calculations from EM1.8/2.0, SV-min, Gogny-HFB, and UNEDF1 are in agreement with the data. $^{37}$Mg meets the separation energy criterion and has become the focus of $1n$ halo nuclei studies in  Mg isotopes.
Turning to  Al isotopes, the discovery of $^{42}$Al suggests a proximity to the drip line, potentially extending to even more neutron-rich isotopes~\cite{Baumann2007}.  Synthesizing theoretical results, we point to $^{40,42,44,46}$Al as satisfying the $S_n$ criterion for a potential halo.
Similar to $S_{n}$, when combining our VS-IMSRG calculations with results from other models for $S_{2n}$, we find that the isotopes of $^{40,42,44,46}$Mg possess the necessary weak binding properties, making them potential candidates for $2n$ halos. Moreover, $^{45,47}$Al, $^{46,48}$Si, $^{49,51}$P, and $^{50,52,54}$S, falling within the shaded area, will be the focus of our study.
Furthermore, the $N=28$ and 32 subshells, present in the calcium chain, are absent in the Mg, Al, Si, P, and S isotopes from the calculated $S_n$ and $S_{2n}$. In addition, the $N=34$ subshell is enhanced in those chains when moving from Ca to Mg isotopes from the predicted $S_{2n}$ using VS-IMSRG.

Merely evaluating small separation energy proves inadequate for confirming halo formation.
Our methodology introduces SFs and TNA as key indicators, complemented by average occupation, drawing insights from established halo-nuclei characteristics. The SF is reflective of the occupation probability of a nucleon within a specific orbital~\cite{TANIHATA2013215}, definition referenced in   Refs.~\cite{XIE2023137800, publisher/Science}.
The SF of $s$ or $p$ wave is typically large in $1n$ halo. Furthermore, the TNA extends this concept to $2n$ halos, which describes the probability amplitude for the simultaneous removal of a pair of nucleons in a specific orbit~\cite{PhysRevC.82.044616}, defined from Ref.~\cite{PhysRevC.74.064604}. 
In the following discussion, we apply the calculated SF and TNA to suggest candidates for $1n$ and $2n$ halo nuclei. 
For a $1n$ halo, the occupation percent of $1n$ in the $s$ or $p$ wave should be larger than that in other types of nuclei. 
A similar situation occurs for the $2n$ halos, which give a large percent of the occupation of $2n$ in the $s$ or $p$ wave.

\begin{figure}[!htb]
\includegraphics[width=1.00\columnwidth]{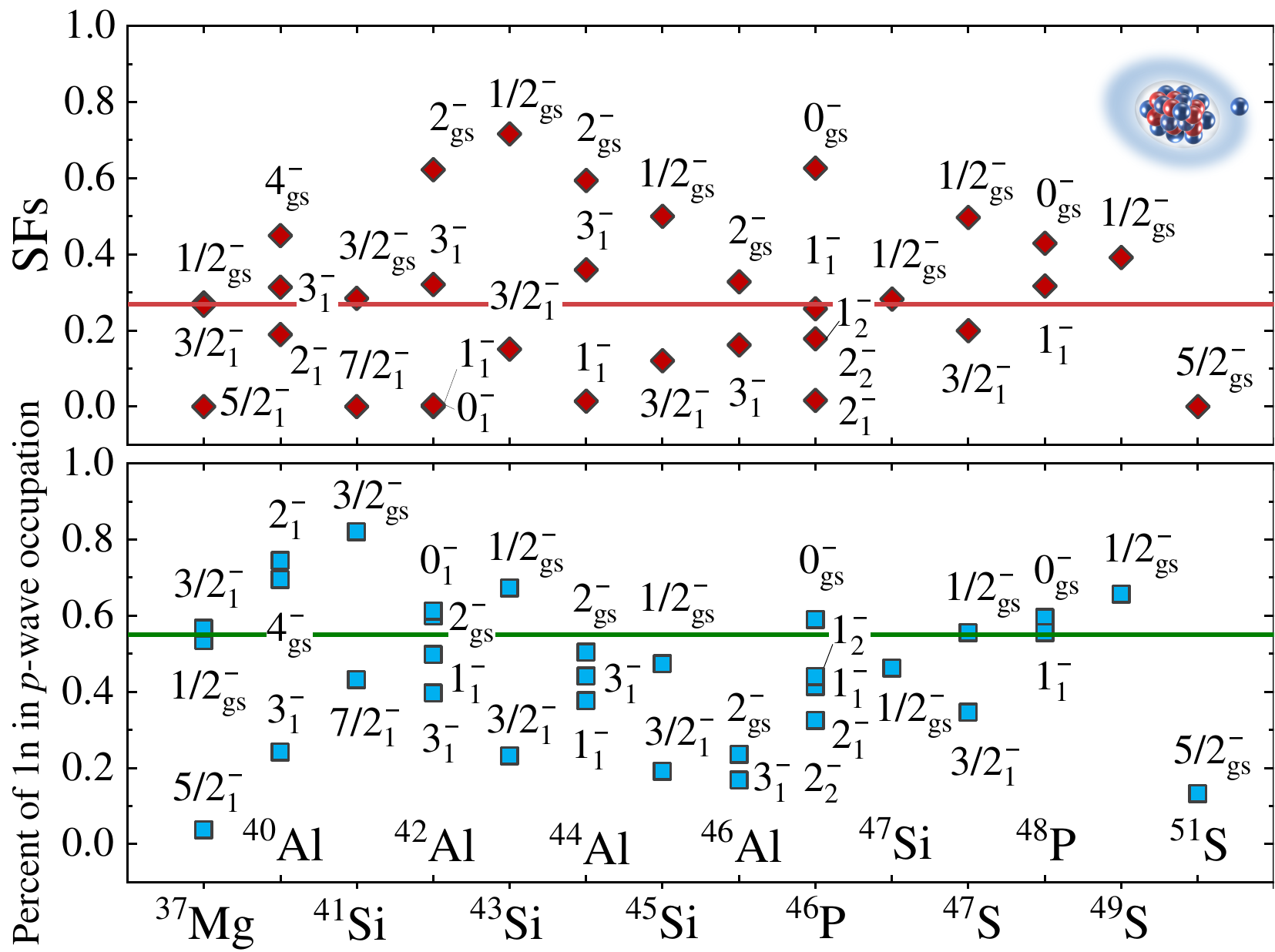}
\vspace{-0.2cm}
\caption{SF and the percent of $1n$ in $p$-wave occupations of nuclei satisfying the $S_n$ condition were calculated using the EM1.8/2.0 interaction. The solid red and green lines represent the results of $^{37}$Mg as the reference.}
\label{SF}
\end{figure}

We first consider the $1n$ halos and calculate the SF and percent of $1n$ in $p$-wave occupancy for $^{37}$Mg, $^{40,42,44,46}$Al, $^{41,43,45,47}$Si, $^{46,48}$P, and $^{47,49,51}$S, presented in the upper and lower panels of Fig.~\ref{SF}, respectively. These isotopes exhibit a small $S_n$, considering only the $p$ waves ($p_{1/2,3/2}$).
Due to the uncertainty of our \textit{ab initio} calculations, the low-lying states with excitation energies ($E_{ex}$) less than 0.5 MeV are considered, and the associated SFs are calculated for the g.s. of the daughter nuclei. The results from  $NN + 3N$(lnl) and EM1.8/2.0 interactions are almost identical. Thus only results of  EM1.8/2.0 are presented in the following.


Combining the experimental data with theoretical shell model calculations, Ref.~\cite{PhysRevLett.112.242501} concluded that the g.s. $^{37}$Mg$_{\rm{g.s.}}$ is a weakly bound $p$-wave halo with a spin-parity of either $3/2^-$ or $1/2^-$.
Our VS-IMSRG calculations give the g.s. of $^{37}$Mg as $1/2^-$ with a closely aligned $3/2^-_1$ state. Moreover, the SFs for the $1/2_{\rm{g.s.}}^-$ and $3/2_1^-$ states are also very close, being 0.272 and 0.267, respectively, corresponding to the significant percentages of $1n$ in $p$-wave occupations. The results align with shell model calculations in Ref.~\cite{PhysRevLett.112.242501}.
Furthermore, the $5/2^-_1$ excited state, with its small SF value, corresponds to a minor percentage of $1n$ in $p$-wave occupation. 
SF and $1n$ occupations both suggest that $3/2_1^-$ and $1/2_1^-$ states possess halo characteristics. 
The appearance of the $p$ orbital in the g.s. of $^{37}$Mg shows a breakdown of the $N=28$ shell gap. The discovery is also crucial in understanding the features of the \textit{island of inversion}.
The results align with the experimental data and the shell model calculations in Ref.~\cite{PhysRevLett.112.242501}, inconsistent with the result from the HFB calculations in Ref.~\cite{PhysRevC.98.011301}.
This alignment of our calculation with experimental measures underscores the validity of our approach in suggesting the halo candidate. 



The average SF of the $1/2_{\rm{g.s.}}^-$ and $3/2_1^-$ states and the $p$-wave average occupation of $1n$ in $^{37}$Mg serves as the reference for $1n$ candidate halo studies, delineated by the solid red and green lines in the upper and lower panels of Fig.~\ref{SF}, respectively.
The large SF of $p$ waves indicates an increased propensity for valence neutrons to occupy the $p_{3/2,1/2}$ orbitals, thereby enhancing the potential for halo formation.
By establishing a threshold SF value above which a nucleus can be deemed a candidate for a $1n$ halo, we suggest the potential candidates for halo structures in the g.s. of $^{40,42,44,46}$Al, $^{41,43,45,47}$Si, $^{46,48}$P, and $^{47,49}$S. We exclude $^{51}$S due to its bearing a small $p$-wave SF.
Drawing on the calculated $1n$ occupancy, we predict a strong possibility of a halo in $^{40,42}$Al, $^{41,43}$Si, $^{46,48}$P, and $^{49}$S. These predictions exhibit a high degree of consistency with those derived from SF calculations. Notwithstanding the slight discrepancies for $^{44}$Al, $^{45}$Si, and $^{47}$Si, the $p$-wave occupancy in these isotopes is remarkably similar to that in $^{37}$Mg, thereby qualifying them as potential halo candidates within our defined margin of error.
Moreover, the excited states, such as $^{40}$Al($3_1^-$), $^{42}$Al($3_1^-$), $^{44}$Al($3_1^-$), and $^{48}$P($1_1^-$), also exhibit traits indicative of a $1n$ halo. 
Noteworthy is the case of $^{42}$Al, currently recognized as the most neutron-rich odd-odd aluminum isotope observed to date~\cite{Baumann2007}. Upcoming experimental work for $^{42}$Al will help validate the \textit{ab initio} predictions.

\begin{figure}[!htb]
\includegraphics[width=0.92\columnwidth]{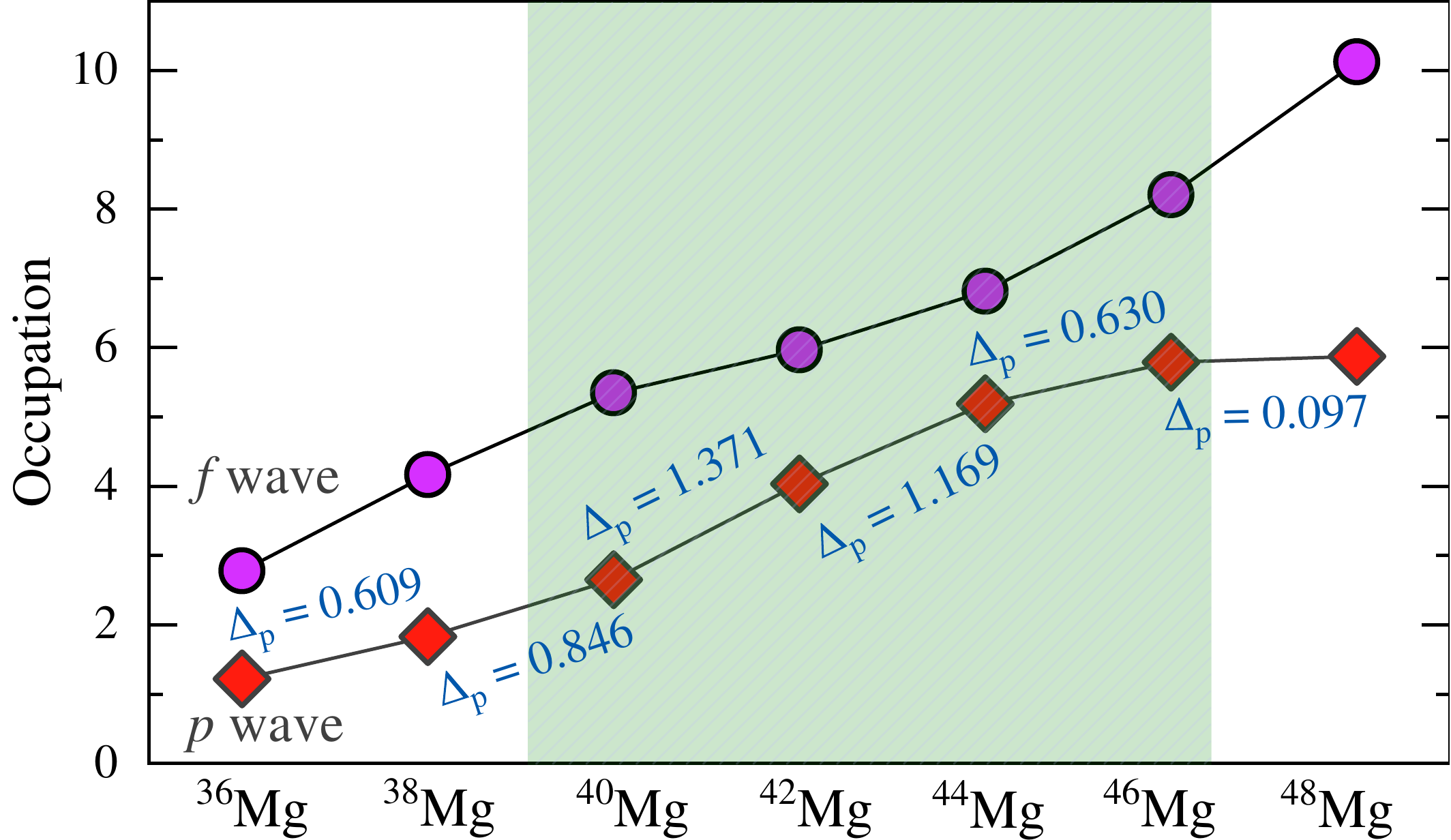}
\vspace{-0.2cm}
\caption{The calculated $p$- and $f$-wave occupations for Mg isotopes using the EM1.8/2.0 interaction. $\Delta_p$ represents the increase in $p$-wave occupation relative to the core and the shaded areas represent possible candidates for $2n$ halos.}
\label{Mg}
\end{figure}

To comprehend $2n$ halos, we first
examine the neutron occupation for Mg isotopes.
Our results, as depicted in Fig.~\ref{Mg}, elucidate the evolution of occupation in $p$ (1$p_{3/2,1/2}$) and $f$ waves (0$f_{7/2,5/2}$). The green-shaded region indicates possible candidates for $2n$ halos, according to ${S_{2n}}$.
Our results reveal a pronounced convex trend in the $p$-wave occupancy as neutron numbers increase.
We also analyzed the occupation of the pair of weakly bound neutrons in the $p$ wave ($\Delta_p$), specifically examining the difference in $p$-wave occupancy between the parent ($A$) and daughter ($A-2$) nuclei.
The results indicate that there is an enhanced $p$-wave occupation in $^{42,44}$Mg. Additionally, the valence $2n$ occupancy in $^{40,46}$Mg is also found to be significant.

\begin{figure}[!htb]
\includegraphics[width=1.00\columnwidth]{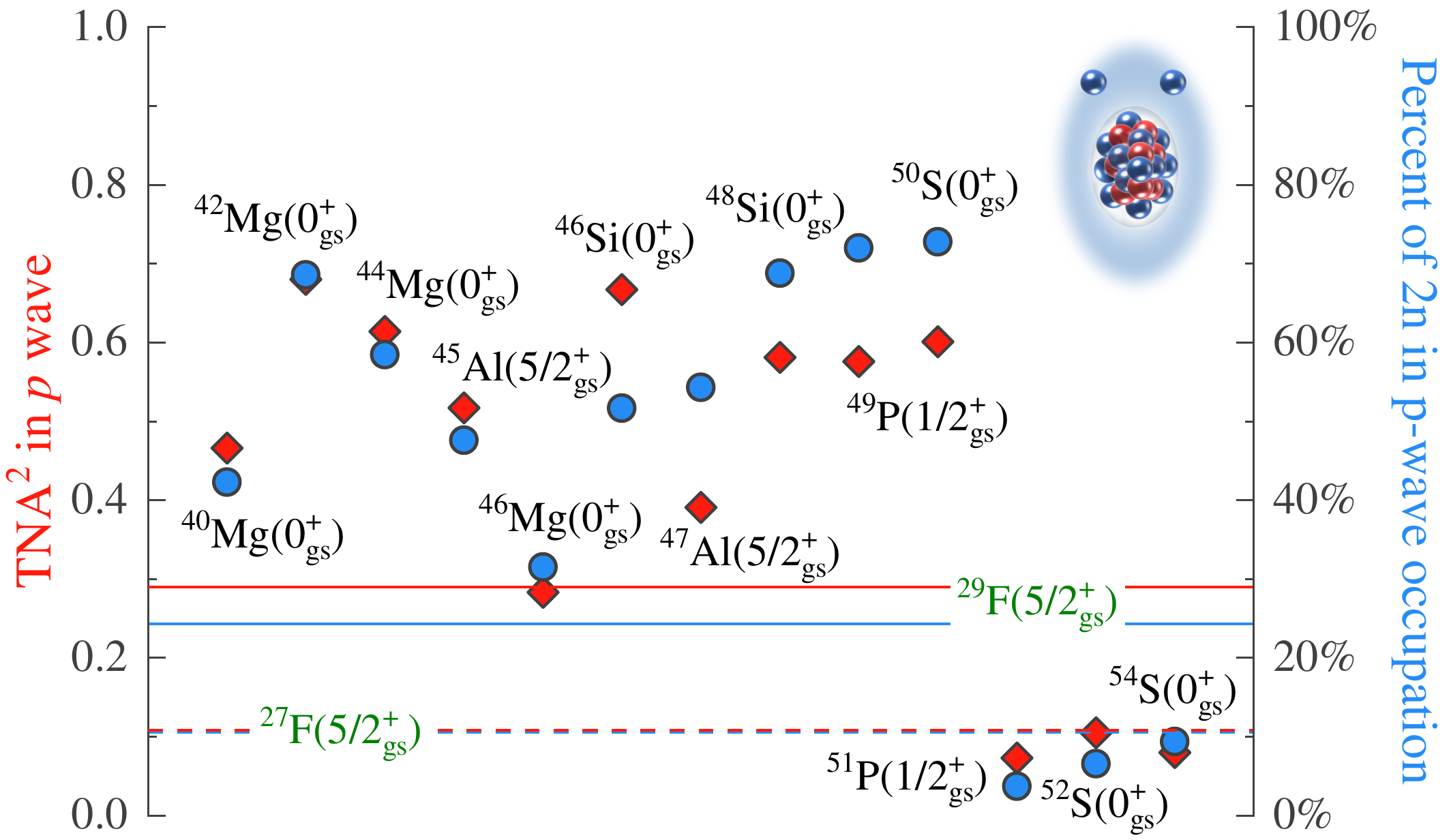}
\vspace{-0.2cm}
\caption{Results of VS-IMSRG calculations with EM1.8/2.0 interactions of TNA$^2$ in the $p$ wave(red rhombus) and the percent of $2n$ in $p$-wave occupation (blue round). The solid line represents the threshold with $^{29}$F as the reference and the dashed line represents $^{27}$F.} 
\label{TNA}
\end{figure}

To further suggest and analyze  potential $2n$ halos, the calculated sum of $\rm{TNA}^2$ for two nucleons occupying the same $p$ orbital, is presented in Fig.~\ref{TNA} (labeled as TNA$^2$ in the $p$ wave), alongside the percentage of $2n$ in the $p$-wave occupancy, i.e., $\Delta_p/2$, for comparison.
Within the TNA calculations, we primarily focus on the spin-parity consistency between the g.s. of the parent nucleus (with $A$ nucleons) and the daughter nucleus (with $A-2$ nucleons), where the transferred neutron pair ($I=0$) is predominant.

\renewcommand{\arraystretch}{1.15}

\begin{table*}[!htb]
\caption{The
relative mean-square neutron radius (in units of fm) between halos and the \textit{inner core} for the suggested $1n$ or $2n$ halo candidates, calculated using the VS-IMSRG with the EM1.8/2.0 interaction. } \label{Rn}
\setlength{\tabcolsep}{0.80mm}{\begin{tabular}{cccccccccccccccccccc}
\hline \hline
Nuclei & \multicolumn{2}{c}{$^{37}$Mg} & \multicolumn{2}{c}{$^{40}$Al} & $^{41}$Si & \multicolumn{2}{c}{$^{42}$Al} & $^{43}$Si &  \multicolumn{2}{c}{$^{44}$Al}  & $^{45}$Si & $^{46}$Al & $^{46}$P &  $^{47}$Si & $^{47}$S & \multicolumn{2}{c}{$^{48}$P} & $^{49}$S  \\ \cline{2-3} \cline{4-5} \cline{7-8} \cline{10-11} \cline{17-18}
 $J^\pi$ &  1/2$^-_{\rm{g.s.}}$ & 3/2$^-_1$ & 4$^-_{\rm{g.s.}}$ & 3$^-_1$ &   3/2$^-_{\rm{g.s.}}$ & 2$^-_{\rm{g.s.}}$ & 3$^-_1$ & 1/2$^-_{\rm{g.s.}}$ &   2$^-_{\rm{g.s.}}$ & 3$^-_1$ & 1/2$^-_{\rm{g.s.}}$ & 2$^-_{\rm{g.s.}}$ & 0$^-_{\rm{g.s.}}$ & 1/2$^-_{\rm{g.s.}}$ &  1/2$^-_{\rm{g.s.}}$ & 0$^-_{\rm{g.s.}}$ & 1$^-_1$ & 1/2$^-_{\rm{g.s.}}$ \\
 $\Delta R_n$& 0.644 & 0.655 & 0.621 & 0.532 & 0.630 & 0.636 & 0.568 & 0.661 & 0.556 & 0.559 & 0.554 & 0.414 & 0.580 & 0.489 & 0.589 & 0.508 & 0.526 & 0.547 \\ \hline

Nuclei & $^{40}$Mg & $^{42}$Mg & $^{44}$Mg &  $^{45}$Al  & $^{46}$Si & $^{47}$Al & $^{48}$Si &  $^{49}$P & $^{50}$S \\
$J^\pi$ &  0$^+_{\rm{gs}}$ &  0$^+_{\rm{gs}}$  &  0$^+_{\rm{gs}}$  &  5/2$^+_{\rm{gs}}$  &     0$^+_{\rm{gs}}$  & 5/2$^+_{\rm{gs}}$ & 0$^+_{\rm{gs}}$  &  1/2$^+_{\rm{gs}}$ & 0$^+_{\rm{gs}}$   \\
$\Delta R_{2n}$ & 0.825 & 0.909 & 0.834 & 0.767 &  0.763 & 0.750 & 0.791 & 0.807 & 0.815  \\
\hline \hline 
\end{tabular}}
\end{table*}

The reference case of $^{29}$F, an experimentally confirmed $2n$ halo nucleus~\cite{PhysRevLett.124.222504}, is calculated using the cross-shell SDPF-MU interactions, as seen by the red and blue solid lines in Fig.~\ref{TNA}.
Moreover, for comparison, $^{27}$F, which is not a halo, is also calculated.
The calculated values of TNA$^2$ in the $p$ wave and the percent of $2n$ in $p$-wave occupation are large in $^{29}$F compared to that of $^{27}$F.
The analysis of $2n$ occupancy percentages and TNA$^2$ in various isotopes yields consistent inferences regarding potential halo nuclei. Specifically, $p$-wave occupation percentages and TNA$^2$ values in the g.s. of $^{51}$P and $^{52,54}$S are akin to those in $^{27}$F, yet fall short of the halo reference set by $^{29}$F. 
This phenomenon is attributed to the appearance of the $N=34$ shell closure in these isotopes, where the valence neutrons in isotopes beyond the $N=34$ subshell are nearly absent in the $p$ wave.
This suggests that these isotopes are not candidates for halo nuclei. Conversely, the g.s. of $^{40,42,44,46}$Mg isotopes 
 demonstrates promising signs as $2n$ halo candidates, echoing our prior occupation predictions.
The predicted $2n$ halos for $^{40,42,44}$Mg align with that of HFB calculations~\cite{PhysRevC.98.011301,PhysRevC.88.054305}. 
Extending this methodology to Al, Si, P, and S isotopes, we have identified potential $2n$ halo candidates, such as the g.s. of $^{45,47}$Al, $^{46,48}$Si, $^{49}$P, and $^{50}$S.

Our approach, incorporating TNA as a tool for forecasting $2n$ halos, offers an accessible observation through experimental means. 
As demonstrated in Fig.~\ref{TNA}, the concurrence between TNA and occupation calculations confirms the reliability of the view. 
Despite some discrepancies, such as those observed in $^{48}$Si, these can be considered negligible when taking into account the variations between TNA and occupation, as demonstrated in the case of $^{29}$F.
However, confirmation of these candidate halo nuclei requires experimental verification.

Finally, we calculate the
relative mean-square neutron radius between halo nuclei and \textit{inner core}, donated by $\Delta R_n$/$\Delta R_{2n}$, which are defined as $\sqrt{\langle R_n^2 \rangle_{A+1/A+2} - \langle R_n^2 \rangle_A}$, using VS-IMSRG calculations based on EM1.8/2.0. The results are shown in the upper and lower panels of Table~\ref{Rn} for suggested $1n$ and $2n$ halo candidates, respectively. The continuum coupling is absent in the present VS-IMSRG calculations, which should enhance the $\Delta R_{n/2n}$ values and reinforce the halo features.


The disappearances of the $N=28$ and 32 subshells in the Mg, Al, Si, P, and S isotopes extend the neutron-dripline in those isotopes and contribute significantly to the existence of $1n$ and $2n$ neutron halos in this region.
This situation is in line with the disappearance of the $N=20$ subshell in fluorine isotopes, which extends the dripline to $^{31}$F \cite{Tsunoda2020} and contributes to the $2n$ halo structure in $^{29}$F. Future experimental studies, including direct mass, SF, and TNA measurements targeting these nuclei, could provide crucial insights into the nature of halo phenomena in medium-mass nuclei, thereby expanding our comprehension of nuclear structures at the extremes of the nuclear landscape.

\textit{Summary.}~
We conducted a comprehensive investigation of halo nuclear phenomena across the isotopic chains of Mg, Al, Si, P, and S using the \textit{ab initio} VS-IMSRG. 
We have accurately replicated the observed odd-even staggering effect in $S_n$ and the decreasing trend with $N$ increasing in $S_{2n}$. This precise depiction allowed us to suggest a series of weakly bound nuclei as potential candidates for further study. We introduced a view to evaluate these nuclei for halo phenomena, leveraging SF and TNA for $1n$ and $2n$, respectively. Using $^{37}$Mg as a reference for $1n$ halo nuclei, the SF calculation shows consistency with the result of the $p$-wave occupation, leading to the prediction that $^{40,42,44,46}$Al, $^{41,43,45,47}$Si, $^{46,48}$P, and $^{47,49}$S are potential $1n$ halos candidates. Similarly, using $^{29}$F as a reference standard for $2n$ halo nuclei, our findings from TNA calculations suggest the g.s. of $^{40,42,44,46}$Mg, $^{45,47}$Al, $^{46,48}$Si, $^{49}$P, and $^{50}$S as promising candidates for $2n$ halos.  Finally, we unveil the interplay between shell evolution and the emergence of halo structures in nuclei, illuminating a fundamental aspect of nuclear physics.
Notably, SF and TNA are both experimentally extractable. Hence, our strategy provides an experimental correlation between the SF and TNA for the existence of $1n$ and $2n$ halos, respectively.


\textit{Acknowledgments.}~
This work has been supported by the National Key R\&D Program of China under Grant No. 2023YFA1606403; the National Natural Science Foundation of China under Grant Nos.  12205340, 12175281, 12347106, and 12121005;  the Gansu Natural Science Foundation under Grant Nos. 22JR5RA123 and 23JRRA614;  the Strategic Priority Research Program of Chinese Academy of Sciences under Grant No. XDB34000000; the Key Research Program of the Chinese Academy of Sciences under Grant No. XDPB15; the State Key Laboratory of Nuclear Physics and Technology, Peking University under Grant No. NPT2020KFY13. The numerical calculations in this paper have been done on Hefei advanced computing center.

\bibliography{Ref}

\begin{thebibliography}{64}%
\makeatletter
\providecommand \@ifxundefined [1]{%
 \@ifx{#1\undefined}
}%
\providecommand \@ifnum [1]{%
 \ifnum #1\expandafter \@firstoftwo
 \else \expandafter \@secondoftwo
 \fi
}%
\providecommand \@ifx [1]{%
 \ifx #1\expandafter \@firstoftwo
 \else \expandafter \@secondoftwo
 \fi
}%
\providecommand \natexlab [1]{#1}%
\providecommand \enquote  [1]{``#1''}%
\providecommand \bibnamefont  [1]{#1}%
\providecommand \bibfnamefont [1]{#1}%
\providecommand \citenamefont [1]{#1}%
\providecommand \href@noop [0]{\@secondoftwo}%
\providecommand \href [0]{\begingroup \@sanitize@url \@href}%
\providecommand \@href[1]{\@@startlink{#1}\@@href}%
\providecommand \@@href[1]{\endgroup#1\@@endlink}%
\providecommand \@sanitize@url [0]{\catcode `\\12\catcode `\$12\catcode `\&12\catcode `\#12\catcode `\^12\catcode `\_12\catcode `\%12\relax}%
\providecommand \@@startlink[1]{}%
\providecommand \@@endlink[0]{}%
\providecommand \url  [0]{\begingroup\@sanitize@url \@url }%
\providecommand \@url [1]{\endgroup\@href {#1}{\urlprefix }}%
\providecommand \urlprefix  [0]{URL }%
\providecommand \Eprint [0]{\href }%
\providecommand \doibase [0]{http://dx.doi.org/}%
\providecommand \selectlanguage [0]{\@gobble}%
\providecommand \bibinfo  [0]{\@secondoftwo}%
\providecommand \bibfield  [0]{\@secondoftwo}%
\providecommand \translation [1]{[#1]}%
\providecommand \BibitemOpen [0]{}%
\providecommand \bibitemStop [0]{}%
\providecommand \bibitemNoStop [0]{.\EOS\space}%
\providecommand \EOS [0]{\spacefactor3000\relax}%
\providecommand \BibitemShut  [1]{\csname bibitem#1\endcsname}%
\let\auto@bib@innerbib\@empty
\bibitem [{\citenamefont {Wang}\ \emph {et~al.}(2004)\citenamefont {Wang}, \citenamefont {Mueller}, \citenamefont {Bailey}, \citenamefont {Drake}, \citenamefont {Greene}, \citenamefont {Henderson}, \citenamefont {Holt}, \citenamefont {Janssens}, \citenamefont {Jiang}, \citenamefont {Lu}, \citenamefont {O'Connor}, \citenamefont {Pardo}, \citenamefont {Rehm}, \citenamefont {Schiffer},\ and\ \citenamefont {Tang}}]{PhysRevLett.93.142501}%
  \BibitemOpen
  \bibfield  {author} {\bibinfo {author} {\bibfnamefont {L.-B.}\ \bibnamefont {Wang}}, \bibinfo {author} {\bibfnamefont {P.}~\bibnamefont {Mueller}}, \bibinfo {author} {\bibfnamefont {K.}~\bibnamefont {Bailey}}, \bibinfo {author} {\bibfnamefont {G.~W.~F.}\ \bibnamefont {Drake}}, \bibinfo {author} {\bibfnamefont {J.~P.}\ \bibnamefont {Greene}}, \bibinfo {author} {\bibfnamefont {D.}~\bibnamefont {Henderson}}, \bibinfo {author} {\bibfnamefont {R.~J.}\ \bibnamefont {Holt}}, \bibinfo {author} {\bibfnamefont {R.~V.~F.}\ \bibnamefont {Janssens}}, \bibinfo {author} {\bibfnamefont {C.~L.}\ \bibnamefont {Jiang}}, \bibinfo {author} {\bibfnamefont {Z.-T.}\ \bibnamefont {Lu}}, \bibinfo {author} {\bibfnamefont {T.~P.}\ \bibnamefont {O'Connor}}, \bibinfo {author} {\bibfnamefont {R.~C.}\ \bibnamefont {Pardo}}, \bibinfo {author} {\bibfnamefont {K.~E.}\ \bibnamefont {Rehm}}, \bibinfo {author} {\bibfnamefont {J.~P.}\ \bibnamefont {Schiffer}}, \ and\ \bibinfo {author} {\bibfnamefont {X.~D.}\ \bibnamefont {Tang}},\ }\href
  {\doibase 10.1103/PhysRevLett.93.142501} {\bibfield  {journal} {\bibinfo  {journal} {Phys. Rev. Lett.}\ }\textbf {\bibinfo {volume} {93}},\ \bibinfo {pages} {142501} (\bibinfo {year} {2004})}\BibitemShut {NoStop}%
\bibitem [{\citenamefont {Brodeur}\ \emph {et~al.}(2012)\citenamefont {Brodeur}, \citenamefont {Brunner}, \citenamefont {Champagne}, \citenamefont {Ettenauer}, \citenamefont {Smith}, \citenamefont {Lapierre}, \citenamefont {Ringle}, \citenamefont {Ryjkov}, \citenamefont {Bacca}, \citenamefont {Delheij}, \citenamefont {Drake}, \citenamefont {Lunney}, \citenamefont {Schwenk},\ and\ \citenamefont {Dilling}}]{PhysRevLett.108.052504}%
  \BibitemOpen
  \bibfield  {author} {\bibinfo {author} {\bibfnamefont {M.}~\bibnamefont {Brodeur}}, \bibinfo {author} {\bibfnamefont {T.}~\bibnamefont {Brunner}}, \bibinfo {author} {\bibfnamefont {C.}~\bibnamefont {Champagne}}, \bibinfo {author} {\bibfnamefont {S.}~\bibnamefont {Ettenauer}}, \bibinfo {author} {\bibfnamefont {M.~J.}\ \bibnamefont {Smith}}, \bibinfo {author} {\bibfnamefont {A.}~\bibnamefont {Lapierre}}, \bibinfo {author} {\bibfnamefont {R.}~\bibnamefont {Ringle}}, \bibinfo {author} {\bibfnamefont {V.~L.}\ \bibnamefont {Ryjkov}}, \bibinfo {author} {\bibfnamefont {S.}~\bibnamefont {Bacca}}, \bibinfo {author} {\bibfnamefont {P.}~\bibnamefont {Delheij}}, \bibinfo {author} {\bibfnamefont {G.~W.~F.}\ \bibnamefont {Drake}}, \bibinfo {author} {\bibfnamefont {D.}~\bibnamefont {Lunney}}, \bibinfo {author} {\bibfnamefont {A.}~\bibnamefont {Schwenk}}, \ and\ \bibinfo {author} {\bibfnamefont {J.}~\bibnamefont {Dilling}},\ }\href {\doibase 10.1103/PhysRevLett.108.052504} {\bibfield  {journal} {\bibinfo  {journal} {Phys.
  Rev. Lett.}\ }\textbf {\bibinfo {volume} {108}},\ \bibinfo {pages} {052504} (\bibinfo {year} {2012})}\BibitemShut {NoStop}%
\bibitem [{\citenamefont {Tanihata}\ \emph {et~al.}(1985)\citenamefont {Tanihata}, \citenamefont {Hamagaki}, \citenamefont {Hashimoto}, \citenamefont {Shida}, \citenamefont {Yoshikawa}, \citenamefont {Sugimoto}, \citenamefont {Yamakawa}, \citenamefont {Kobayashi},\ and\ \citenamefont {Takahashi}}]{PhysRevLett.55.2676}%
  \BibitemOpen
  \bibfield  {author} {\bibinfo {author} {\bibfnamefont {I.}~\bibnamefont {Tanihata}}, \bibinfo {author} {\bibfnamefont {H.}~\bibnamefont {Hamagaki}}, \bibinfo {author} {\bibfnamefont {O.}~\bibnamefont {Hashimoto}}, \bibinfo {author} {\bibfnamefont {Y.}~\bibnamefont {Shida}}, \bibinfo {author} {\bibfnamefont {N.}~\bibnamefont {Yoshikawa}}, \bibinfo {author} {\bibfnamefont {K.}~\bibnamefont {Sugimoto}}, \bibinfo {author} {\bibfnamefont {O.}~\bibnamefont {Yamakawa}}, \bibinfo {author} {\bibfnamefont {T.}~\bibnamefont {Kobayashi}}, \ and\ \bibinfo {author} {\bibfnamefont {N.}~\bibnamefont {Takahashi}},\ }\href {\doibase 10.1103/PhysRevLett.55.2676} {\bibfield  {journal} {\bibinfo  {journal} {Phys. Rev. Lett.}\ }\textbf {\bibinfo {volume} {55}},\ \bibinfo {pages} {2676} (\bibinfo {year} {1985})}\BibitemShut {NoStop}%
\bibitem [{\citenamefont {Marqu\'es}\ \emph {et~al.}(2001)\citenamefont {Marqu\'es}, \citenamefont {Labiche}, \citenamefont {Orr}, \citenamefont {Ang\'elique}, \citenamefont {Axelsson}, \citenamefont {Benoit}, \citenamefont {Bergmann}, \citenamefont {Borge}, \citenamefont {Catford}, \citenamefont {Chappell}, \citenamefont {Clarke}, \citenamefont {Costa}, \citenamefont {Curtis}, \citenamefont {D'Arrigo}, \citenamefont {Brennand}, \citenamefont {Santos}, \citenamefont {Dorvaux}, \citenamefont {Fazio}, \citenamefont {Freer}, \citenamefont {Fulton}, \citenamefont {Giardina}, \citenamefont {Gr\'evy}, \citenamefont {Guillemaud-Mueller}, \citenamefont {Hanappe}, \citenamefont {Heusch}, \citenamefont {Jonson}, \citenamefont {Brun}, \citenamefont {Leenhardt}, \citenamefont {Lewitowicz}, \citenamefont {L\'opez}, \citenamefont {Markenroth}, \citenamefont {Mueller}, \citenamefont {Nilsson}, \citenamefont {Ninane}, \citenamefont {Nyman}, \citenamefont {Piqueras}, \citenamefont {Riisager}, \citenamefont {Saint~Laurent},
  \citenamefont {Sarazin}, \citenamefont {Singer}, \citenamefont {Sorlin},\ and\ \citenamefont {Stuttg\'e}}]{PhysRevC.64.061301}%
  \BibitemOpen
  \bibfield  {author} {\bibinfo {author} {\bibfnamefont {F.~M.}\ \bibnamefont {Marqu\'es}}, \bibinfo {author} {\bibfnamefont {M.}~\bibnamefont {Labiche}}, \bibinfo {author} {\bibfnamefont {N.~A.}\ \bibnamefont {Orr}}, \bibinfo {author} {\bibfnamefont {J.~C.}\ \bibnamefont {Ang\'elique}}, \bibinfo {author} {\bibfnamefont {L.}~\bibnamefont {Axelsson}}, \bibinfo {author} {\bibfnamefont {B.}~\bibnamefont {Benoit}}, \bibinfo {author} {\bibfnamefont {U.~C.}\ \bibnamefont {Bergmann}}, \bibinfo {author} {\bibfnamefont {M.~J.~G.}\ \bibnamefont {Borge}}, \bibinfo {author} {\bibfnamefont {W.~N.}\ \bibnamefont {Catford}}, \bibinfo {author} {\bibfnamefont {S.~P.~G.}\ \bibnamefont {Chappell}}, \bibinfo {author} {\bibfnamefont {N.~M.}\ \bibnamefont {Clarke}}, \bibinfo {author} {\bibfnamefont {G.}~\bibnamefont {Costa}}, \bibinfo {author} {\bibfnamefont {N.}~\bibnamefont {Curtis}}, \bibinfo {author} {\bibfnamefont {A.}~\bibnamefont {D'Arrigo}}, \bibinfo {author} {\bibfnamefont {E.~d.~G.}\ \bibnamefont {Brennand}}, \bibinfo
  {author} {\bibfnamefont {F.~d.~O.}\ \bibnamefont {Santos}}, \bibinfo {author} {\bibfnamefont {O.}~\bibnamefont {Dorvaux}}, \bibinfo {author} {\bibfnamefont {G.}~\bibnamefont {Fazio}}, \bibinfo {author} {\bibfnamefont {M.}~\bibnamefont {Freer}}, \bibinfo {author} {\bibfnamefont {B.~R.}\ \bibnamefont {Fulton}}, \bibinfo {author} {\bibfnamefont {G.}~\bibnamefont {Giardina}}, \bibinfo {author} {\bibfnamefont {S.}~\bibnamefont {Gr\'evy}}, \bibinfo {author} {\bibfnamefont {D.}~\bibnamefont {Guillemaud-Mueller}}, \bibinfo {author} {\bibfnamefont {F.}~\bibnamefont {Hanappe}}, \bibinfo {author} {\bibfnamefont {B.}~\bibnamefont {Heusch}}, \bibinfo {author} {\bibfnamefont {B.}~\bibnamefont {Jonson}}, \bibinfo {author} {\bibfnamefont {C.~L.}\ \bibnamefont {Brun}}, \bibinfo {author} {\bibfnamefont {S.}~\bibnamefont {Leenhardt}}, \bibinfo {author} {\bibfnamefont {M.}~\bibnamefont {Lewitowicz}}, \bibinfo {author} {\bibfnamefont {M.~J.}\ \bibnamefont {L\'opez}}, \bibinfo {author} {\bibfnamefont {K.}~\bibnamefont
  {Markenroth}}, \bibinfo {author} {\bibfnamefont {A.~C.}\ \bibnamefont {Mueller}}, \bibinfo {author} {\bibfnamefont {T.}~\bibnamefont {Nilsson}}, \bibinfo {author} {\bibfnamefont {A.}~\bibnamefont {Ninane}}, \bibinfo {author} {\bibfnamefont {G.}~\bibnamefont {Nyman}}, \bibinfo {author} {\bibfnamefont {I.}~\bibnamefont {Piqueras}}, \bibinfo {author} {\bibfnamefont {K.}~\bibnamefont {Riisager}}, \bibinfo {author} {\bibfnamefont {M.~G.}\ \bibnamefont {Saint~Laurent}}, \bibinfo {author} {\bibfnamefont {F.}~\bibnamefont {Sarazin}}, \bibinfo {author} {\bibfnamefont {S.~M.}\ \bibnamefont {Singer}}, \bibinfo {author} {\bibfnamefont {O.}~\bibnamefont {Sorlin}}, \ and\ \bibinfo {author} {\bibfnamefont {L.}~\bibnamefont {Stuttg\'e}},\ }\href {\doibase 10.1103/PhysRevC.64.061301} {\bibfield  {journal} {\bibinfo  {journal} {Phys. Rev. C}\ }\textbf {\bibinfo {volume} {64}},\ \bibinfo {pages} {061301(R)} (\bibinfo {year} {2001})}\BibitemShut {NoStop}%
\bibitem [{\citenamefont {Bowman}\ \emph {et~al.}(1973)\citenamefont {Bowman}, \citenamefont {Poskanzer}, \citenamefont {Korteling},\ and\ \citenamefont {Butler}}]{PhysRevLett.31.614}%
  \BibitemOpen
  \bibfield  {author} {\bibinfo {author} {\bibfnamefont {J.~D.}\ \bibnamefont {Bowman}}, \bibinfo {author} {\bibfnamefont {A.~M.}\ \bibnamefont {Poskanzer}}, \bibinfo {author} {\bibfnamefont {R.~G.}\ \bibnamefont {Korteling}}, \ and\ \bibinfo {author} {\bibfnamefont {G.~W.}\ \bibnamefont {Butler}},\ }\href {\doibase 10.1103/PhysRevLett.31.614} {\bibfield  {journal} {\bibinfo  {journal} {Phys. Rev. Lett.}\ }\textbf {\bibinfo {volume} {31}},\ \bibinfo {pages} {614} (\bibinfo {year} {1973})}\BibitemShut {NoStop}%
\bibitem [{\citenamefont {Tanaka}\ \emph {et~al.}(2010)\citenamefont {Tanaka}, \citenamefont {Yamaguchi}, \citenamefont {Suzuki}, \citenamefont {Ohtsubo}, \citenamefont {Fukuda}, \citenamefont {Nishimura}, \citenamefont {Takechi}, \citenamefont {Ogata}, \citenamefont {Ozawa}, \citenamefont {Izumikawa}, \citenamefont {Aiba}, \citenamefont {Aoi}, \citenamefont {Baba}, \citenamefont {Hashizume}, \citenamefont {Inafuku}, \citenamefont {Iwasa}, \citenamefont {Kobayashi}, \citenamefont {Komuro}, \citenamefont {Kondo}, \citenamefont {Kubo}, \citenamefont {Kurokawa}, \citenamefont {Matsuyama}, \citenamefont {Michimasa}, \citenamefont {Motobayashi}, \citenamefont {Nakabayashi}, \citenamefont {Nakajima}, \citenamefont {Nakamura}, \citenamefont {Sakurai}, \citenamefont {Shinoda}, \citenamefont {Shinohara}, \citenamefont {Suzuki}, \citenamefont {Takeshita}, \citenamefont {Takeuchi}, \citenamefont {Togano}, \citenamefont {Yamada}, \citenamefont {Yasuno},\ and\ \citenamefont {Yoshitake}}]{PhysRevLett.104.062701}%
  \BibitemOpen
  \bibfield  {author} {\bibinfo {author} {\bibfnamefont {K.}~\bibnamefont {Tanaka}}, \bibinfo {author} {\bibfnamefont {T.}~\bibnamefont {Yamaguchi}}, \bibinfo {author} {\bibfnamefont {T.}~\bibnamefont {Suzuki}}, \bibinfo {author} {\bibfnamefont {T.}~\bibnamefont {Ohtsubo}}, \bibinfo {author} {\bibfnamefont {M.}~\bibnamefont {Fukuda}}, \bibinfo {author} {\bibfnamefont {D.}~\bibnamefont {Nishimura}}, \bibinfo {author} {\bibfnamefont {M.}~\bibnamefont {Takechi}}, \bibinfo {author} {\bibfnamefont {K.}~\bibnamefont {Ogata}}, \bibinfo {author} {\bibfnamefont {A.}~\bibnamefont {Ozawa}}, \bibinfo {author} {\bibfnamefont {T.}~\bibnamefont {Izumikawa}}, \bibinfo {author} {\bibfnamefont {T.}~\bibnamefont {Aiba}}, \bibinfo {author} {\bibfnamefont {N.}~\bibnamefont {Aoi}}, \bibinfo {author} {\bibfnamefont {H.}~\bibnamefont {Baba}}, \bibinfo {author} {\bibfnamefont {Y.}~\bibnamefont {Hashizume}}, \bibinfo {author} {\bibfnamefont {K.}~\bibnamefont {Inafuku}}, \bibinfo {author} {\bibfnamefont {N.}~\bibnamefont {Iwasa}},
  \bibinfo {author} {\bibfnamefont {K.}~\bibnamefont {Kobayashi}}, \bibinfo {author} {\bibfnamefont {M.}~\bibnamefont {Komuro}}, \bibinfo {author} {\bibfnamefont {Y.}~\bibnamefont {Kondo}}, \bibinfo {author} {\bibfnamefont {T.}~\bibnamefont {Kubo}}, \bibinfo {author} {\bibfnamefont {M.}~\bibnamefont {Kurokawa}}, \bibinfo {author} {\bibfnamefont {T.}~\bibnamefont {Matsuyama}}, \bibinfo {author} {\bibfnamefont {S.}~\bibnamefont {Michimasa}}, \bibinfo {author} {\bibfnamefont {T.}~\bibnamefont {Motobayashi}}, \bibinfo {author} {\bibfnamefont {T.}~\bibnamefont {Nakabayashi}}, \bibinfo {author} {\bibfnamefont {S.}~\bibnamefont {Nakajima}}, \bibinfo {author} {\bibfnamefont {T.}~\bibnamefont {Nakamura}}, \bibinfo {author} {\bibfnamefont {H.}~\bibnamefont {Sakurai}}, \bibinfo {author} {\bibfnamefont {R.}~\bibnamefont {Shinoda}}, \bibinfo {author} {\bibfnamefont {M.}~\bibnamefont {Shinohara}}, \bibinfo {author} {\bibfnamefont {H.}~\bibnamefont {Suzuki}}, \bibinfo {author} {\bibfnamefont {E.}~\bibnamefont {Takeshita}},
  \bibinfo {author} {\bibfnamefont {S.}~\bibnamefont {Takeuchi}}, \bibinfo {author} {\bibfnamefont {Y.}~\bibnamefont {Togano}}, \bibinfo {author} {\bibfnamefont {K.}~\bibnamefont {Yamada}}, \bibinfo {author} {\bibfnamefont {T.}~\bibnamefont {Yasuno}}, \ and\ \bibinfo {author} {\bibfnamefont {M.}~\bibnamefont {Yoshitake}},\ }\href {\doibase 10.1103/PhysRevLett.104.062701} {\bibfield  {journal} {\bibinfo  {journal} {Phys. Rev. Lett.}\ }\textbf {\bibinfo {volume} {104}},\ \bibinfo {pages} {062701} (\bibinfo {year} {2010})}\BibitemShut {NoStop}%
\bibitem [{\citenamefont {Bagchi}\ \emph {et~al.}(2020)\citenamefont {Bagchi}, \citenamefont {Kanungo}, \citenamefont {Tanaka}, \citenamefont {Geissel}, \citenamefont {Doornenbal}, \citenamefont {Horiuchi}, \citenamefont {Hagen}, \citenamefont {Suzuki}, \citenamefont {Tsunoda}, \citenamefont {Ahn}, \citenamefont {Baba}, \citenamefont {Behr}, \citenamefont {Browne}, \citenamefont {Chen}, \citenamefont {Cort\'es}, \citenamefont {Estrad\'e}, \citenamefont {Fukuda}, \citenamefont {Holl}, \citenamefont {Itahashi}, \citenamefont {Iwasa}, \citenamefont {Jansen}, \citenamefont {Jiang}, \citenamefont {Kaur}, \citenamefont {Macchiavelli}, \citenamefont {Matsumoto}, \citenamefont {Momiyama}, \citenamefont {Murray}, \citenamefont {Nakamura}, \citenamefont {Novario}, \citenamefont {Ong}, \citenamefont {Otsuka}, \citenamefont {Papenbrock}, \citenamefont {Paschalis}, \citenamefont {Prochazka}, \citenamefont {Scheidenberger}, \citenamefont {Schrock}, \citenamefont {Shimizu}, \citenamefont {Steppenbeck}, \citenamefont
  {Sakurai}, \citenamefont {Suzuki}, \citenamefont {Suzuki}, \citenamefont {Takechi}, \citenamefont {Takeda}, \citenamefont {Takeuchi}, \citenamefont {Taniuchi}, \citenamefont {Wimmer},\ and\ \citenamefont {Yoshida}}]{PhysRevLett.124.222504}%
  \BibitemOpen
  \bibfield  {author} {\bibinfo {author} {\bibfnamefont {S.}~\bibnamefont {Bagchi}}, \bibinfo {author} {\bibfnamefont {R.}~\bibnamefont {Kanungo}}, \bibinfo {author} {\bibfnamefont {Y.~K.}\ \bibnamefont {Tanaka}}, \bibinfo {author} {\bibfnamefont {H.}~\bibnamefont {Geissel}}, \bibinfo {author} {\bibfnamefont {P.}~\bibnamefont {Doornenbal}}, \bibinfo {author} {\bibfnamefont {W.}~\bibnamefont {Horiuchi}}, \bibinfo {author} {\bibfnamefont {G.}~\bibnamefont {Hagen}}, \bibinfo {author} {\bibfnamefont {T.}~\bibnamefont {Suzuki}}, \bibinfo {author} {\bibfnamefont {N.}~\bibnamefont {Tsunoda}}, \bibinfo {author} {\bibfnamefont {D.~S.}\ \bibnamefont {Ahn}}, \bibinfo {author} {\bibfnamefont {H.}~\bibnamefont {Baba}}, \bibinfo {author} {\bibfnamefont {K.}~\bibnamefont {Behr}}, \bibinfo {author} {\bibfnamefont {F.}~\bibnamefont {Browne}}, \bibinfo {author} {\bibfnamefont {S.}~\bibnamefont {Chen}}, \bibinfo {author} {\bibfnamefont {M.~L.}\ \bibnamefont {Cort\'es}}, \bibinfo {author} {\bibfnamefont {A.}~\bibnamefont
  {Estrad\'e}}, \bibinfo {author} {\bibfnamefont {N.}~\bibnamefont {Fukuda}}, \bibinfo {author} {\bibfnamefont {M.}~\bibnamefont {Holl}}, \bibinfo {author} {\bibfnamefont {K.}~\bibnamefont {Itahashi}}, \bibinfo {author} {\bibfnamefont {N.}~\bibnamefont {Iwasa}}, \bibinfo {author} {\bibfnamefont {G.~R.}\ \bibnamefont {Jansen}}, \bibinfo {author} {\bibfnamefont {W.~G.}\ \bibnamefont {Jiang}}, \bibinfo {author} {\bibfnamefont {S.}~\bibnamefont {Kaur}}, \bibinfo {author} {\bibfnamefont {A.~O.}\ \bibnamefont {Macchiavelli}}, \bibinfo {author} {\bibfnamefont {S.~Y.}\ \bibnamefont {Matsumoto}}, \bibinfo {author} {\bibfnamefont {S.}~\bibnamefont {Momiyama}}, \bibinfo {author} {\bibfnamefont {I.}~\bibnamefont {Murray}}, \bibinfo {author} {\bibfnamefont {T.}~\bibnamefont {Nakamura}}, \bibinfo {author} {\bibfnamefont {S.~J.}\ \bibnamefont {Novario}}, \bibinfo {author} {\bibfnamefont {H.~J.}\ \bibnamefont {Ong}}, \bibinfo {author} {\bibfnamefont {T.}~\bibnamefont {Otsuka}}, \bibinfo {author} {\bibfnamefont
  {T.}~\bibnamefont {Papenbrock}}, \bibinfo {author} {\bibfnamefont {S.}~\bibnamefont {Paschalis}}, \bibinfo {author} {\bibfnamefont {A.}~\bibnamefont {Prochazka}}, \bibinfo {author} {\bibfnamefont {C.}~\bibnamefont {Scheidenberger}}, \bibinfo {author} {\bibfnamefont {P.}~\bibnamefont {Schrock}}, \bibinfo {author} {\bibfnamefont {Y.}~\bibnamefont {Shimizu}}, \bibinfo {author} {\bibfnamefont {D.}~\bibnamefont {Steppenbeck}}, \bibinfo {author} {\bibfnamefont {H.}~\bibnamefont {Sakurai}}, \bibinfo {author} {\bibfnamefont {D.}~\bibnamefont {Suzuki}}, \bibinfo {author} {\bibfnamefont {H.}~\bibnamefont {Suzuki}}, \bibinfo {author} {\bibfnamefont {M.}~\bibnamefont {Takechi}}, \bibinfo {author} {\bibfnamefont {H.}~\bibnamefont {Takeda}}, \bibinfo {author} {\bibfnamefont {S.}~\bibnamefont {Takeuchi}}, \bibinfo {author} {\bibfnamefont {R.}~\bibnamefont {Taniuchi}}, \bibinfo {author} {\bibfnamefont {K.}~\bibnamefont {Wimmer}}, \ and\ \bibinfo {author} {\bibfnamefont {K.}~\bibnamefont {Yoshida}},\ }\href {\doibase
  10.1103/PhysRevLett.124.222504} {\bibfield  {journal} {\bibinfo  {journal} {Phys. Rev. Lett.}\ }\textbf {\bibinfo {volume} {124}},\ \bibinfo {pages} {222504} (\bibinfo {year} {2020})}\BibitemShut {NoStop}%
\bibitem [{\citenamefont {Kelley}\ \emph {et~al.}(1995)\citenamefont {Kelley}, \citenamefont {Austin}, \citenamefont {Kryger}, \citenamefont {Morrissey}, \citenamefont {Orr}, \citenamefont {Sherrill}, \citenamefont {Thoennessen}, \citenamefont {Winfield}, \citenamefont {Winger},\ and\ \citenamefont {Young}}]{PhysRevLett.74.30}%
  \BibitemOpen
  \bibfield  {author} {\bibinfo {author} {\bibfnamefont {J.~H.}\ \bibnamefont {Kelley}}, \bibinfo {author} {\bibfnamefont {S.~M.}\ \bibnamefont {Austin}}, \bibinfo {author} {\bibfnamefont {R.~A.}\ \bibnamefont {Kryger}}, \bibinfo {author} {\bibfnamefont {D.~J.}\ \bibnamefont {Morrissey}}, \bibinfo {author} {\bibfnamefont {N.~A.}\ \bibnamefont {Orr}}, \bibinfo {author} {\bibfnamefont {B.~M.}\ \bibnamefont {Sherrill}}, \bibinfo {author} {\bibfnamefont {M.}~\bibnamefont {Thoennessen}}, \bibinfo {author} {\bibfnamefont {J.~S.}\ \bibnamefont {Winfield}}, \bibinfo {author} {\bibfnamefont {J.~A.}\ \bibnamefont {Winger}}, \ and\ \bibinfo {author} {\bibfnamefont {B.~M.}\ \bibnamefont {Young}},\ }\href {\doibase 10.1103/PhysRevLett.74.30} {\bibfield  {journal} {\bibinfo  {journal} {Phys. Rev. Lett.}\ }\textbf {\bibinfo {volume} {74}},\ \bibinfo {pages} {30} (\bibinfo {year} {1995})}\BibitemShut {NoStop}%
\bibitem [{\citenamefont {Fang}\ \emph {et~al.}(2004)\citenamefont {Fang}, \citenamefont {Yamaguchi}, \citenamefont {Zheng}, \citenamefont {Ozawa}, \citenamefont {Chiba}, \citenamefont {Kanungo}, \citenamefont {Kato}, \citenamefont {Morimoto}, \citenamefont {Ohnishi}, \citenamefont {Suda}, \citenamefont {Yamaguchi}, \citenamefont {Yoshida}, \citenamefont {Yoshida},\ and\ \citenamefont {Tanihata}}]{PhysRevC.69.034613}%
  \BibitemOpen
  \bibfield  {author} {\bibinfo {author} {\bibfnamefont {D.~Q.}\ \bibnamefont {Fang}}, \bibinfo {author} {\bibfnamefont {T.}~\bibnamefont {Yamaguchi}}, \bibinfo {author} {\bibfnamefont {T.}~\bibnamefont {Zheng}}, \bibinfo {author} {\bibfnamefont {A.}~\bibnamefont {Ozawa}}, \bibinfo {author} {\bibfnamefont {M.}~\bibnamefont {Chiba}}, \bibinfo {author} {\bibfnamefont {R.}~\bibnamefont {Kanungo}}, \bibinfo {author} {\bibfnamefont {T.}~\bibnamefont {Kato}}, \bibinfo {author} {\bibfnamefont {K.}~\bibnamefont {Morimoto}}, \bibinfo {author} {\bibfnamefont {T.}~\bibnamefont {Ohnishi}}, \bibinfo {author} {\bibfnamefont {T.}~\bibnamefont {Suda}}, \bibinfo {author} {\bibfnamefont {Y.}~\bibnamefont {Yamaguchi}}, \bibinfo {author} {\bibfnamefont {A.}~\bibnamefont {Yoshida}}, \bibinfo {author} {\bibfnamefont {K.}~\bibnamefont {Yoshida}}, \ and\ \bibinfo {author} {\bibfnamefont {I.}~\bibnamefont {Tanihata}},\ }\href {\doibase 10.1103/PhysRevC.69.034613} {\bibfield  {journal} {\bibinfo  {journal} {Phys. Rev. C}\ }\textbf
  {\bibinfo {volume} {69}},\ \bibinfo {pages} {034613} (\bibinfo {year} {2004})}\BibitemShut {NoStop}%
\bibitem [{\citenamefont {Alcorta}\ \emph {et~al.}(2011)\citenamefont {Alcorta}, \citenamefont {Rehm}, \citenamefont {Back}, \citenamefont {Bedoor}, \citenamefont {Bertone}, \citenamefont {Deibel}, \citenamefont {DiGiovine}, \citenamefont {Esbensen}, \citenamefont {Greene}, \citenamefont {Hoffman}, \citenamefont {Jiang}, \citenamefont {Lighthall}, \citenamefont {Marley}, \citenamefont {Pardo}, \citenamefont {Paul}, \citenamefont {Rogers}, \citenamefont {Ugalde},\ and\ \citenamefont {Wuosmaa}}]{PhysRevLett.106.172701}%
  \BibitemOpen
  \bibfield  {author} {\bibinfo {author} {\bibfnamefont {M.}~\bibnamefont {Alcorta}}, \bibinfo {author} {\bibfnamefont {K.~E.}\ \bibnamefont {Rehm}}, \bibinfo {author} {\bibfnamefont {B.~B.}\ \bibnamefont {Back}}, \bibinfo {author} {\bibfnamefont {S.}~\bibnamefont {Bedoor}}, \bibinfo {author} {\bibfnamefont {P.~F.}\ \bibnamefont {Bertone}}, \bibinfo {author} {\bibfnamefont {C.~M.}\ \bibnamefont {Deibel}}, \bibinfo {author} {\bibfnamefont {B.}~\bibnamefont {DiGiovine}}, \bibinfo {author} {\bibfnamefont {H.}~\bibnamefont {Esbensen}}, \bibinfo {author} {\bibfnamefont {J.~P.}\ \bibnamefont {Greene}}, \bibinfo {author} {\bibfnamefont {C.~R.}\ \bibnamefont {Hoffman}}, \bibinfo {author} {\bibfnamefont {C.~L.}\ \bibnamefont {Jiang}}, \bibinfo {author} {\bibfnamefont {J.~C.}\ \bibnamefont {Lighthall}}, \bibinfo {author} {\bibfnamefont {S.~T.}\ \bibnamefont {Marley}}, \bibinfo {author} {\bibfnamefont {R.~C.}\ \bibnamefont {Pardo}}, \bibinfo {author} {\bibfnamefont {M.}~\bibnamefont {Paul}}, \bibinfo {author}
  {\bibfnamefont {A.~M.}\ \bibnamefont {Rogers}}, \bibinfo {author} {\bibfnamefont {C.}~\bibnamefont {Ugalde}}, \ and\ \bibinfo {author} {\bibfnamefont {A.~H.}\ \bibnamefont {Wuosmaa}},\ }\href {\doibase 10.1103/PhysRevLett.106.172701} {\bibfield  {journal} {\bibinfo  {journal} {Phys. Rev. Lett.}\ }\textbf {\bibinfo {volume} {106}},\ \bibinfo {pages} {172701} (\bibinfo {year} {2011})}\BibitemShut {NoStop}%
\bibitem [{\citenamefont {Bazin}\ \emph {et~al.}(1995)\citenamefont {Bazin}, \citenamefont {Brown}, \citenamefont {Brown}, \citenamefont {Fauerbach}, \citenamefont {Hellstr\"om}, \citenamefont {Hirzebruch}, \citenamefont {Kelley}, \citenamefont {Kryger}, \citenamefont {Morrissey}, \citenamefont {Pfaff}, \citenamefont {Powell}, \citenamefont {Sherrill},\ and\ \citenamefont {Thoennessen}}]{PhysRevLett.74.3569}%
  \BibitemOpen
  \bibfield  {author} {\bibinfo {author} {\bibfnamefont {D.}~\bibnamefont {Bazin}}, \bibinfo {author} {\bibfnamefont {B.~A.}\ \bibnamefont {Brown}}, \bibinfo {author} {\bibfnamefont {J.}~\bibnamefont {Brown}}, \bibinfo {author} {\bibfnamefont {M.}~\bibnamefont {Fauerbach}}, \bibinfo {author} {\bibfnamefont {M.}~\bibnamefont {Hellstr\"om}}, \bibinfo {author} {\bibfnamefont {S.~E.}\ \bibnamefont {Hirzebruch}}, \bibinfo {author} {\bibfnamefont {J.~H.}\ \bibnamefont {Kelley}}, \bibinfo {author} {\bibfnamefont {R.~A.}\ \bibnamefont {Kryger}}, \bibinfo {author} {\bibfnamefont {D.~J.}\ \bibnamefont {Morrissey}}, \bibinfo {author} {\bibfnamefont {R.}~\bibnamefont {Pfaff}}, \bibinfo {author} {\bibfnamefont {C.~F.}\ \bibnamefont {Powell}}, \bibinfo {author} {\bibfnamefont {B.~M.}\ \bibnamefont {Sherrill}}, \ and\ \bibinfo {author} {\bibfnamefont {M.}~\bibnamefont {Thoennessen}},\ }\href {\doibase 10.1103/PhysRevLett.74.3569} {\bibfield  {journal} {\bibinfo  {journal} {Phys. Rev. Lett.}\ }\textbf {\bibinfo {volume}
  {74}},\ \bibinfo {pages} {3569} (\bibinfo {year} {1995})}\BibitemShut {NoStop}%
\bibitem [{\citenamefont {Nakamura}\ \emph {et~al.}(1999)\citenamefont {Nakamura}, \citenamefont {Fukuda}, \citenamefont {Kobayashi}, \citenamefont {Aoi}, \citenamefont {Iwasaki}, \citenamefont {Kubo}, \citenamefont {Mengoni}, \citenamefont {Notani}, \citenamefont {Otsu}, \citenamefont {Sakurai}, \citenamefont {Shimoura}, \citenamefont {Teranishi}, \citenamefont {Watanabe}, \citenamefont {Yoneda},\ and\ \citenamefont {Ishihara}}]{PhysRevLett.83.1112}%
  \BibitemOpen
  \bibfield  {author} {\bibinfo {author} {\bibfnamefont {T.}~\bibnamefont {Nakamura}}, \bibinfo {author} {\bibfnamefont {N.}~\bibnamefont {Fukuda}}, \bibinfo {author} {\bibfnamefont {T.}~\bibnamefont {Kobayashi}}, \bibinfo {author} {\bibfnamefont {N.}~\bibnamefont {Aoi}}, \bibinfo {author} {\bibfnamefont {H.}~\bibnamefont {Iwasaki}}, \bibinfo {author} {\bibfnamefont {T.}~\bibnamefont {Kubo}}, \bibinfo {author} {\bibfnamefont {A.}~\bibnamefont {Mengoni}}, \bibinfo {author} {\bibfnamefont {M.}~\bibnamefont {Notani}}, \bibinfo {author} {\bibfnamefont {H.}~\bibnamefont {Otsu}}, \bibinfo {author} {\bibfnamefont {H.}~\bibnamefont {Sakurai}}, \bibinfo {author} {\bibfnamefont {S.}~\bibnamefont {Shimoura}}, \bibinfo {author} {\bibfnamefont {T.}~\bibnamefont {Teranishi}}, \bibinfo {author} {\bibfnamefont {Y.~X.}\ \bibnamefont {Watanabe}}, \bibinfo {author} {\bibfnamefont {K.}~\bibnamefont {Yoneda}}, \ and\ \bibinfo {author} {\bibfnamefont {M.}~\bibnamefont {Ishihara}},\ }\href {\doibase 10.1103/PhysRevLett.83.1112}
  {\bibfield  {journal} {\bibinfo  {journal} {Phys. Rev. Lett.}\ }\textbf {\bibinfo {volume} {83}},\ \bibinfo {pages} {1112} (\bibinfo {year} {1999})}\BibitemShut {NoStop}%
\bibitem [{\citenamefont {Nakamura}\ \emph {et~al.}(2014)\citenamefont {Nakamura}, \citenamefont {Kobayashi}, \citenamefont {Kondo}, \citenamefont {Satou}, \citenamefont {Tostevin}, \citenamefont {Utsuno}, \citenamefont {Aoi}, \citenamefont {Baba}, \citenamefont {Fukuda}, \citenamefont {Gibelin}, \citenamefont {Inabe}, \citenamefont {Ishihara}, \citenamefont {Kameda}, \citenamefont {Kubo}, \citenamefont {Motobayashi}, \citenamefont {Ohnishi}, \citenamefont {Orr}, \citenamefont {Otsu}, \citenamefont {Otsuka}, \citenamefont {Sakurai}, \citenamefont {Sumikama}, \citenamefont {Takeda}, \citenamefont {Takeshita}, \citenamefont {Takechi}, \citenamefont {Takeuchi}, \citenamefont {Togano},\ and\ \citenamefont {Yoneda}}]{PhysRevLett.112.142501}%
  \BibitemOpen
  \bibfield  {author} {\bibinfo {author} {\bibfnamefont {T.}~\bibnamefont {Nakamura}}, \bibinfo {author} {\bibfnamefont {N.}~\bibnamefont {Kobayashi}}, \bibinfo {author} {\bibfnamefont {Y.}~\bibnamefont {Kondo}}, \bibinfo {author} {\bibfnamefont {Y.}~\bibnamefont {Satou}}, \bibinfo {author} {\bibfnamefont {J.~A.}\ \bibnamefont {Tostevin}}, \bibinfo {author} {\bibfnamefont {Y.}~\bibnamefont {Utsuno}}, \bibinfo {author} {\bibfnamefont {N.}~\bibnamefont {Aoi}}, \bibinfo {author} {\bibfnamefont {H.}~\bibnamefont {Baba}}, \bibinfo {author} {\bibfnamefont {N.}~\bibnamefont {Fukuda}}, \bibinfo {author} {\bibfnamefont {J.}~\bibnamefont {Gibelin}}, \bibinfo {author} {\bibfnamefont {N.}~\bibnamefont {Inabe}}, \bibinfo {author} {\bibfnamefont {M.}~\bibnamefont {Ishihara}}, \bibinfo {author} {\bibfnamefont {D.}~\bibnamefont {Kameda}}, \bibinfo {author} {\bibfnamefont {T.}~\bibnamefont {Kubo}}, \bibinfo {author} {\bibfnamefont {T.}~\bibnamefont {Motobayashi}}, \bibinfo {author} {\bibfnamefont {T.}~\bibnamefont {Ohnishi}},
  \bibinfo {author} {\bibfnamefont {N.~A.}\ \bibnamefont {Orr}}, \bibinfo {author} {\bibfnamefont {H.}~\bibnamefont {Otsu}}, \bibinfo {author} {\bibfnamefont {T.}~\bibnamefont {Otsuka}}, \bibinfo {author} {\bibfnamefont {H.}~\bibnamefont {Sakurai}}, \bibinfo {author} {\bibfnamefont {T.}~\bibnamefont {Sumikama}}, \bibinfo {author} {\bibfnamefont {H.}~\bibnamefont {Takeda}}, \bibinfo {author} {\bibfnamefont {E.}~\bibnamefont {Takeshita}}, \bibinfo {author} {\bibfnamefont {M.}~\bibnamefont {Takechi}}, \bibinfo {author} {\bibfnamefont {S.}~\bibnamefont {Takeuchi}}, \bibinfo {author} {\bibfnamefont {Y.}~\bibnamefont {Togano}}, \ and\ \bibinfo {author} {\bibfnamefont {K.}~\bibnamefont {Yoneda}},\ }\href {\doibase 10.1103/PhysRevLett.112.142501} {\bibfield  {journal} {\bibinfo  {journal} {Phys. Rev. Lett.}\ }\textbf {\bibinfo {volume} {112}},\ \bibinfo {pages} {142501} (\bibinfo {year} {2014})}\BibitemShut {NoStop}%
\bibitem [{\citenamefont {Kobayashi}\ \emph {et~al.}(2014)\citenamefont {Kobayashi}, \citenamefont {Nakamura}, \citenamefont {Kondo}, \citenamefont {Tostevin}, \citenamefont {Utsuno}, \citenamefont {Aoi}, \citenamefont {Baba}, \citenamefont {Barthelemy}, \citenamefont {Famiano}, \citenamefont {Fukuda}, \citenamefont {Inabe}, \citenamefont {Ishihara}, \citenamefont {Kanungo}, \citenamefont {Kim}, \citenamefont {Kubo}, \citenamefont {Lee}, \citenamefont {Lee}, \citenamefont {Matsushita}, \citenamefont {Motobayashi}, \citenamefont {Ohnishi}, \citenamefont {Orr}, \citenamefont {Otsu}, \citenamefont {Otsuka}, \citenamefont {Sako}, \citenamefont {Sakurai}, \citenamefont {Satou}, \citenamefont {Sumikama}, \citenamefont {Takeda}, \citenamefont {Takeuchi}, \citenamefont {Tanaka}, \citenamefont {Togano},\ and\ \citenamefont {Yoneda}}]{PhysRevLett.112.242501}%
  \BibitemOpen
  \bibfield  {author} {\bibinfo {author} {\bibfnamefont {N.}~\bibnamefont {Kobayashi}}, \bibinfo {author} {\bibfnamefont {T.}~\bibnamefont {Nakamura}}, \bibinfo {author} {\bibfnamefont {Y.}~\bibnamefont {Kondo}}, \bibinfo {author} {\bibfnamefont {J.~A.}\ \bibnamefont {Tostevin}}, \bibinfo {author} {\bibfnamefont {Y.}~\bibnamefont {Utsuno}}, \bibinfo {author} {\bibfnamefont {N.}~\bibnamefont {Aoi}}, \bibinfo {author} {\bibfnamefont {H.}~\bibnamefont {Baba}}, \bibinfo {author} {\bibfnamefont {R.}~\bibnamefont {Barthelemy}}, \bibinfo {author} {\bibfnamefont {M.~A.}\ \bibnamefont {Famiano}}, \bibinfo {author} {\bibfnamefont {N.}~\bibnamefont {Fukuda}}, \bibinfo {author} {\bibfnamefont {N.}~\bibnamefont {Inabe}}, \bibinfo {author} {\bibfnamefont {M.}~\bibnamefont {Ishihara}}, \bibinfo {author} {\bibfnamefont {R.}~\bibnamefont {Kanungo}}, \bibinfo {author} {\bibfnamefont {S.}~\bibnamefont {Kim}}, \bibinfo {author} {\bibfnamefont {T.}~\bibnamefont {Kubo}}, \bibinfo {author} {\bibfnamefont {G.~S.}\ \bibnamefont
  {Lee}}, \bibinfo {author} {\bibfnamefont {H.~S.}\ \bibnamefont {Lee}}, \bibinfo {author} {\bibfnamefont {M.}~\bibnamefont {Matsushita}}, \bibinfo {author} {\bibfnamefont {T.}~\bibnamefont {Motobayashi}}, \bibinfo {author} {\bibfnamefont {T.}~\bibnamefont {Ohnishi}}, \bibinfo {author} {\bibfnamefont {N.~A.}\ \bibnamefont {Orr}}, \bibinfo {author} {\bibfnamefont {H.}~\bibnamefont {Otsu}}, \bibinfo {author} {\bibfnamefont {T.}~\bibnamefont {Otsuka}}, \bibinfo {author} {\bibfnamefont {T.}~\bibnamefont {Sako}}, \bibinfo {author} {\bibfnamefont {H.}~\bibnamefont {Sakurai}}, \bibinfo {author} {\bibfnamefont {Y.}~\bibnamefont {Satou}}, \bibinfo {author} {\bibfnamefont {T.}~\bibnamefont {Sumikama}}, \bibinfo {author} {\bibfnamefont {H.}~\bibnamefont {Takeda}}, \bibinfo {author} {\bibfnamefont {S.}~\bibnamefont {Takeuchi}}, \bibinfo {author} {\bibfnamefont {R.}~\bibnamefont {Tanaka}}, \bibinfo {author} {\bibfnamefont {Y.}~\bibnamefont {Togano}}, \ and\ \bibinfo {author} {\bibfnamefont {K.}~\bibnamefont {Yoneda}},\
  }\href {\doibase 10.1103/PhysRevLett.112.242501} {\bibfield  {journal} {\bibinfo  {journal} {Phys. Rev. Lett.}\ }\textbf {\bibinfo {volume} {112}},\ \bibinfo {pages} {242501} (\bibinfo {year} {2014})}\BibitemShut {NoStop}%
\bibitem [{\citenamefont {Tanihata}\ \emph {et~al.}(2013)\citenamefont {Tanihata}, \citenamefont {Savajols},\ and\ \citenamefont {Kanungo}}]{TANIHATA2013215}%
  \BibitemOpen
  \bibfield  {author} {\bibinfo {author} {\bibfnamefont {I.}~\bibnamefont {Tanihata}}, \bibinfo {author} {\bibfnamefont {H.}~\bibnamefont {Savajols}}, \ and\ \bibinfo {author} {\bibfnamefont {R.}~\bibnamefont {Kanungo}},\ }\href {\doibase https://doi.org/10.1016/j.ppnp.2012.07.001} {\bibfield  {journal} {\bibinfo  {journal} {Progress in Particle and Nuclear Physics}\ }\textbf {\bibinfo {volume} {68}},\ \bibinfo {pages} {215} (\bibinfo {year} {2013})}\BibitemShut {NoStop}%
\bibitem [{\citenamefont {Takechi}\ \emph {et~al.}(2014)\citenamefont {Takechi}, \citenamefont {Suzuki}, \citenamefont {Nishimura}, \citenamefont {Fukuda}, \citenamefont {Ohtsubo}, \citenamefont {Nagashima}, \citenamefont {Suzuki}, \citenamefont {Yamaguchi}, \citenamefont {Ozawa}, \citenamefont {Moriguchi}, \citenamefont {Ohishi}, \citenamefont {Sumikama}, \citenamefont {Geissel}, \citenamefont {Aoi}, \citenamefont {Chen}, \citenamefont {Fang}, \citenamefont {Fukuda}, \citenamefont {Fukuoka}, \citenamefont {Furuki}, \citenamefont {Inabe}, \citenamefont {Ishibashi}, \citenamefont {Itoh}, \citenamefont {Izumikawa}, \citenamefont {Kameda}, \citenamefont {Kubo}, \citenamefont {Lantz}, \citenamefont {Lee}, \citenamefont {Ma}, \citenamefont {Matsuta}, \citenamefont {Mihara}, \citenamefont {Momota}, \citenamefont {Nagae}, \citenamefont {Nishikiori}, \citenamefont {Niwa}, \citenamefont {Ohnishi}, \citenamefont {Okumura}, \citenamefont {Ohtake}, \citenamefont {Ogura}, \citenamefont {Sakurai}, \citenamefont {Sato},
  \citenamefont {Shimbara}, \citenamefont {Suzuki}, \citenamefont {Takeda}, \citenamefont {Takeuchi}, \citenamefont {Tanaka}, \citenamefont {Tanaka}, \citenamefont {Uenishi}, \citenamefont {Winkler}, \citenamefont {Yanagisawa}, \citenamefont {Watanabe}, \citenamefont {Minomo}, \citenamefont {Tagami}, \citenamefont {Shimada}, \citenamefont {Kimura}, \citenamefont {Matsumoto}, \citenamefont {Shimizu},\ and\ \citenamefont {Yahiro}}]{PhysRevC.90.061305}%
  \BibitemOpen
  \bibfield  {author} {\bibinfo {author} {\bibfnamefont {M.}~\bibnamefont {Takechi}}, \bibinfo {author} {\bibfnamefont {S.}~\bibnamefont {Suzuki}}, \bibinfo {author} {\bibfnamefont {D.}~\bibnamefont {Nishimura}}, \bibinfo {author} {\bibfnamefont {M.}~\bibnamefont {Fukuda}}, \bibinfo {author} {\bibfnamefont {T.}~\bibnamefont {Ohtsubo}}, \bibinfo {author} {\bibfnamefont {M.}~\bibnamefont {Nagashima}}, \bibinfo {author} {\bibfnamefont {T.}~\bibnamefont {Suzuki}}, \bibinfo {author} {\bibfnamefont {T.}~\bibnamefont {Yamaguchi}}, \bibinfo {author} {\bibfnamefont {A.}~\bibnamefont {Ozawa}}, \bibinfo {author} {\bibfnamefont {T.}~\bibnamefont {Moriguchi}}, \bibinfo {author} {\bibfnamefont {H.}~\bibnamefont {Ohishi}}, \bibinfo {author} {\bibfnamefont {T.}~\bibnamefont {Sumikama}}, \bibinfo {author} {\bibfnamefont {H.}~\bibnamefont {Geissel}}, \bibinfo {author} {\bibfnamefont {N.}~\bibnamefont {Aoi}}, \bibinfo {author} {\bibfnamefont {R.-J.}\ \bibnamefont {Chen}}, \bibinfo {author} {\bibfnamefont {D.-Q.}\ \bibnamefont
  {Fang}}, \bibinfo {author} {\bibfnamefont {N.}~\bibnamefont {Fukuda}}, \bibinfo {author} {\bibfnamefont {S.}~\bibnamefont {Fukuoka}}, \bibinfo {author} {\bibfnamefont {H.}~\bibnamefont {Furuki}}, \bibinfo {author} {\bibfnamefont {N.}~\bibnamefont {Inabe}}, \bibinfo {author} {\bibfnamefont {Y.}~\bibnamefont {Ishibashi}}, \bibinfo {author} {\bibfnamefont {T.}~\bibnamefont {Itoh}}, \bibinfo {author} {\bibfnamefont {T.}~\bibnamefont {Izumikawa}}, \bibinfo {author} {\bibfnamefont {D.}~\bibnamefont {Kameda}}, \bibinfo {author} {\bibfnamefont {T.}~\bibnamefont {Kubo}}, \bibinfo {author} {\bibfnamefont {M.}~\bibnamefont {Lantz}}, \bibinfo {author} {\bibfnamefont {C.~S.}\ \bibnamefont {Lee}}, \bibinfo {author} {\bibfnamefont {Y.-G.}\ \bibnamefont {Ma}}, \bibinfo {author} {\bibfnamefont {K.}~\bibnamefont {Matsuta}}, \bibinfo {author} {\bibfnamefont {M.}~\bibnamefont {Mihara}}, \bibinfo {author} {\bibfnamefont {S.}~\bibnamefont {Momota}}, \bibinfo {author} {\bibfnamefont {D.}~\bibnamefont {Nagae}}, \bibinfo {author}
  {\bibfnamefont {R.}~\bibnamefont {Nishikiori}}, \bibinfo {author} {\bibfnamefont {T.}~\bibnamefont {Niwa}}, \bibinfo {author} {\bibfnamefont {T.}~\bibnamefont {Ohnishi}}, \bibinfo {author} {\bibfnamefont {K.}~\bibnamefont {Okumura}}, \bibinfo {author} {\bibfnamefont {M.}~\bibnamefont {Ohtake}}, \bibinfo {author} {\bibfnamefont {T.}~\bibnamefont {Ogura}}, \bibinfo {author} {\bibfnamefont {H.}~\bibnamefont {Sakurai}}, \bibinfo {author} {\bibfnamefont {K.}~\bibnamefont {Sato}}, \bibinfo {author} {\bibfnamefont {Y.}~\bibnamefont {Shimbara}}, \bibinfo {author} {\bibfnamefont {H.}~\bibnamefont {Suzuki}}, \bibinfo {author} {\bibfnamefont {H.}~\bibnamefont {Takeda}}, \bibinfo {author} {\bibfnamefont {S.}~\bibnamefont {Takeuchi}}, \bibinfo {author} {\bibfnamefont {K.}~\bibnamefont {Tanaka}}, \bibinfo {author} {\bibfnamefont {M.}~\bibnamefont {Tanaka}}, \bibinfo {author} {\bibfnamefont {H.}~\bibnamefont {Uenishi}}, \bibinfo {author} {\bibfnamefont {M.}~\bibnamefont {Winkler}}, \bibinfo {author} {\bibfnamefont
  {Y.}~\bibnamefont {Yanagisawa}}, \bibinfo {author} {\bibfnamefont {S.}~\bibnamefont {Watanabe}}, \bibinfo {author} {\bibfnamefont {K.}~\bibnamefont {Minomo}}, \bibinfo {author} {\bibfnamefont {S.}~\bibnamefont {Tagami}}, \bibinfo {author} {\bibfnamefont {M.}~\bibnamefont {Shimada}}, \bibinfo {author} {\bibfnamefont {M.}~\bibnamefont {Kimura}}, \bibinfo {author} {\bibfnamefont {T.}~\bibnamefont {Matsumoto}}, \bibinfo {author} {\bibfnamefont {Y.~R.}\ \bibnamefont {Shimizu}}, \ and\ \bibinfo {author} {\bibfnamefont {M.}~\bibnamefont {Yahiro}},\ }\href {\doibase 10.1103/PhysRevC.90.061305} {\bibfield  {journal} {\bibinfo  {journal} {Phys. Rev. C}\ }\textbf {\bibinfo {volume} {90}},\ \bibinfo {pages} {061305(R)} (\bibinfo {year} {2014})}\BibitemShut {NoStop}%
\bibitem [{\citenamefont {Nakamura}\ \emph {et~al.}(2009)\citenamefont {Nakamura}, \citenamefont {Kobayashi}, \citenamefont {Kondo}, \citenamefont {Satou}, \citenamefont {Aoi}, \citenamefont {Baba}, \citenamefont {Deguchi}, \citenamefont {Fukuda}, \citenamefont {Gibelin}, \citenamefont {Inabe}, \citenamefont {Ishihara}, \citenamefont {Kameda}, \citenamefont {Kawada}, \citenamefont {Kubo}, \citenamefont {Kusaka}, \citenamefont {Mengoni}, \citenamefont {Motobayashi}, \citenamefont {Ohnishi}, \citenamefont {Ohtake}, \citenamefont {Orr}, \citenamefont {Otsu}, \citenamefont {Otsuka}, \citenamefont {Saito}, \citenamefont {Sakurai}, \citenamefont {Shimoura}, \citenamefont {Sumikama}, \citenamefont {Takeda}, \citenamefont {Takeshita}, \citenamefont {Takechi}, \citenamefont {Takeuchi}, \citenamefont {Tanaka}, \citenamefont {Tanaka}, \citenamefont {Tanaka}, \citenamefont {Togano}, \citenamefont {Utsuno}, \citenamefont {Yoneda}, \citenamefont {Yoshida},\ and\ \citenamefont {Yoshida}}]{PhysRevLett.103.262501}%
  \BibitemOpen
  \bibfield  {author} {\bibinfo {author} {\bibfnamefont {T.}~\bibnamefont {Nakamura}}, \bibinfo {author} {\bibfnamefont {N.}~\bibnamefont {Kobayashi}}, \bibinfo {author} {\bibfnamefont {Y.}~\bibnamefont {Kondo}}, \bibinfo {author} {\bibfnamefont {Y.}~\bibnamefont {Satou}}, \bibinfo {author} {\bibfnamefont {N.}~\bibnamefont {Aoi}}, \bibinfo {author} {\bibfnamefont {H.}~\bibnamefont {Baba}}, \bibinfo {author} {\bibfnamefont {S.}~\bibnamefont {Deguchi}}, \bibinfo {author} {\bibfnamefont {N.}~\bibnamefont {Fukuda}}, \bibinfo {author} {\bibfnamefont {J.}~\bibnamefont {Gibelin}}, \bibinfo {author} {\bibfnamefont {N.}~\bibnamefont {Inabe}}, \bibinfo {author} {\bibfnamefont {M.}~\bibnamefont {Ishihara}}, \bibinfo {author} {\bibfnamefont {D.}~\bibnamefont {Kameda}}, \bibinfo {author} {\bibfnamefont {Y.}~\bibnamefont {Kawada}}, \bibinfo {author} {\bibfnamefont {T.}~\bibnamefont {Kubo}}, \bibinfo {author} {\bibfnamefont {K.}~\bibnamefont {Kusaka}}, \bibinfo {author} {\bibfnamefont {A.}~\bibnamefont {Mengoni}}, \bibinfo
  {author} {\bibfnamefont {T.}~\bibnamefont {Motobayashi}}, \bibinfo {author} {\bibfnamefont {T.}~\bibnamefont {Ohnishi}}, \bibinfo {author} {\bibfnamefont {M.}~\bibnamefont {Ohtake}}, \bibinfo {author} {\bibfnamefont {N.~A.}\ \bibnamefont {Orr}}, \bibinfo {author} {\bibfnamefont {H.}~\bibnamefont {Otsu}}, \bibinfo {author} {\bibfnamefont {T.}~\bibnamefont {Otsuka}}, \bibinfo {author} {\bibfnamefont {A.}~\bibnamefont {Saito}}, \bibinfo {author} {\bibfnamefont {H.}~\bibnamefont {Sakurai}}, \bibinfo {author} {\bibfnamefont {S.}~\bibnamefont {Shimoura}}, \bibinfo {author} {\bibfnamefont {T.}~\bibnamefont {Sumikama}}, \bibinfo {author} {\bibfnamefont {H.}~\bibnamefont {Takeda}}, \bibinfo {author} {\bibfnamefont {E.}~\bibnamefont {Takeshita}}, \bibinfo {author} {\bibfnamefont {M.}~\bibnamefont {Takechi}}, \bibinfo {author} {\bibfnamefont {S.}~\bibnamefont {Takeuchi}}, \bibinfo {author} {\bibfnamefont {K.}~\bibnamefont {Tanaka}}, \bibinfo {author} {\bibfnamefont {K.~N.}\ \bibnamefont {Tanaka}}, \bibinfo {author}
  {\bibfnamefont {N.}~\bibnamefont {Tanaka}}, \bibinfo {author} {\bibfnamefont {Y.}~\bibnamefont {Togano}}, \bibinfo {author} {\bibfnamefont {Y.}~\bibnamefont {Utsuno}}, \bibinfo {author} {\bibfnamefont {K.}~\bibnamefont {Yoneda}}, \bibinfo {author} {\bibfnamefont {A.}~\bibnamefont {Yoshida}}, \ and\ \bibinfo {author} {\bibfnamefont {K.}~\bibnamefont {Yoshida}},\ }\href {\doibase 10.1103/PhysRevLett.103.262501} {\bibfield  {journal} {\bibinfo  {journal} {Phys. Rev. Lett.}\ }\textbf {\bibinfo {volume} {103}},\ \bibinfo {pages} {262501} (\bibinfo {year} {2009})}\BibitemShut {NoStop}%
\bibitem [{\citenamefont {Simon}\ \emph {et~al.}(1999)\citenamefont {Simon}, \citenamefont {Aleksandrov}, \citenamefont {Aumann}, \citenamefont {Axelsson}, \citenamefont {Baumann}, \citenamefont {Borge}, \citenamefont {Chulkov}, \citenamefont {Collatz}, \citenamefont {Cub}, \citenamefont {Dostal}, \citenamefont {Eberlein}, \citenamefont {Elze}, \citenamefont {Emling}, \citenamefont {Geissel}, \citenamefont {Gr\"unschloss}, \citenamefont {Hellstr\"om}, \citenamefont {Holeczek}, \citenamefont {Holzmann}, \citenamefont {Jonson}, \citenamefont {Kratz}, \citenamefont {Kraus}, \citenamefont {Kulessa}, \citenamefont {Leifels}, \citenamefont {Leistenschneider}, \citenamefont {Leth}, \citenamefont {Mukha}, \citenamefont {M\"unzenberg}, \citenamefont {Nickel}, \citenamefont {Nilsson}, \citenamefont {Nyman}, \citenamefont {Petersen}, \citenamefont {Pf\"utzner}, \citenamefont {Richter}, \citenamefont {Riisager}, \citenamefont {Scheidenberger}, \citenamefont {Schrieder}, \citenamefont {Schwab}, \citenamefont {Smedberg},
  \citenamefont {Stroth}, \citenamefont {Surowiec}, \citenamefont {Tengblad},\ and\ \citenamefont {Zhukov}}]{PhysRevLett.83.496}%
  \BibitemOpen
  \bibfield  {author} {\bibinfo {author} {\bibfnamefont {H.}~\bibnamefont {Simon}}, \bibinfo {author} {\bibfnamefont {D.}~\bibnamefont {Aleksandrov}}, \bibinfo {author} {\bibfnamefont {T.}~\bibnamefont {Aumann}}, \bibinfo {author} {\bibfnamefont {L.}~\bibnamefont {Axelsson}}, \bibinfo {author} {\bibfnamefont {T.}~\bibnamefont {Baumann}}, \bibinfo {author} {\bibfnamefont {M.~J.~G.}\ \bibnamefont {Borge}}, \bibinfo {author} {\bibfnamefont {L.~V.}\ \bibnamefont {Chulkov}}, \bibinfo {author} {\bibfnamefont {R.}~\bibnamefont {Collatz}}, \bibinfo {author} {\bibfnamefont {J.}~\bibnamefont {Cub}}, \bibinfo {author} {\bibfnamefont {W.}~\bibnamefont {Dostal}}, \bibinfo {author} {\bibfnamefont {B.}~\bibnamefont {Eberlein}}, \bibinfo {author} {\bibfnamefont {T.~W.}\ \bibnamefont {Elze}}, \bibinfo {author} {\bibfnamefont {H.}~\bibnamefont {Emling}}, \bibinfo {author} {\bibfnamefont {H.}~\bibnamefont {Geissel}}, \bibinfo {author} {\bibfnamefont {A.}~\bibnamefont {Gr\"unschloss}}, \bibinfo {author} {\bibfnamefont
  {M.}~\bibnamefont {Hellstr\"om}}, \bibinfo {author} {\bibfnamefont {J.}~\bibnamefont {Holeczek}}, \bibinfo {author} {\bibfnamefont {R.}~\bibnamefont {Holzmann}}, \bibinfo {author} {\bibfnamefont {B.}~\bibnamefont {Jonson}}, \bibinfo {author} {\bibfnamefont {J.~V.}\ \bibnamefont {Kratz}}, \bibinfo {author} {\bibfnamefont {G.}~\bibnamefont {Kraus}}, \bibinfo {author} {\bibfnamefont {R.}~\bibnamefont {Kulessa}}, \bibinfo {author} {\bibfnamefont {Y.}~\bibnamefont {Leifels}}, \bibinfo {author} {\bibfnamefont {A.}~\bibnamefont {Leistenschneider}}, \bibinfo {author} {\bibfnamefont {T.}~\bibnamefont {Leth}}, \bibinfo {author} {\bibfnamefont {I.}~\bibnamefont {Mukha}}, \bibinfo {author} {\bibfnamefont {G.}~\bibnamefont {M\"unzenberg}}, \bibinfo {author} {\bibfnamefont {F.}~\bibnamefont {Nickel}}, \bibinfo {author} {\bibfnamefont {T.}~\bibnamefont {Nilsson}}, \bibinfo {author} {\bibfnamefont {G.}~\bibnamefont {Nyman}}, \bibinfo {author} {\bibfnamefont {B.}~\bibnamefont {Petersen}}, \bibinfo {author} {\bibfnamefont
  {M.}~\bibnamefont {Pf\"utzner}}, \bibinfo {author} {\bibfnamefont {A.}~\bibnamefont {Richter}}, \bibinfo {author} {\bibfnamefont {K.}~\bibnamefont {Riisager}}, \bibinfo {author} {\bibfnamefont {C.}~\bibnamefont {Scheidenberger}}, \bibinfo {author} {\bibfnamefont {G.}~\bibnamefont {Schrieder}}, \bibinfo {author} {\bibfnamefont {W.}~\bibnamefont {Schwab}}, \bibinfo {author} {\bibfnamefont {M.~H.}\ \bibnamefont {Smedberg}}, \bibinfo {author} {\bibfnamefont {J.}~\bibnamefont {Stroth}}, \bibinfo {author} {\bibfnamefont {A.}~\bibnamefont {Surowiec}}, \bibinfo {author} {\bibfnamefont {O.}~\bibnamefont {Tengblad}}, \ and\ \bibinfo {author} {\bibfnamefont {M.~V.}\ \bibnamefont {Zhukov}},\ }\href {\doibase 10.1103/PhysRevLett.83.496} {\bibfield  {journal} {\bibinfo  {journal} {Phys. Rev. Lett.}\ }\textbf {\bibinfo {volume} {83}},\ \bibinfo {pages} {496} (\bibinfo {year} {1999})}\BibitemShut {NoStop}%
\bibitem [{\citenamefont {Tanihata}\ \emph {et~al.}(2008)\citenamefont {Tanihata}, \citenamefont {Alcorta}, \citenamefont {Bandyopadhyay}, \citenamefont {Bieri}, \citenamefont {Buchmann}, \citenamefont {Davids}, \citenamefont {Galinski}, \citenamefont {Howell}, \citenamefont {Mills}, \citenamefont {Mythili}, \citenamefont {Openshaw}, \citenamefont {Padilla-Rodal}, \citenamefont {Ruprecht}, \citenamefont {Sheffer}, \citenamefont {Shotter}, \citenamefont {Trinczek}, \citenamefont {Walden}, \citenamefont {Savajols}, \citenamefont {Roger}, \citenamefont {Caamano}, \citenamefont {Mittig}, \citenamefont {Roussel-Chomaz}, \citenamefont {Kanungo}, \citenamefont {Gallant}, \citenamefont {Notani}, \citenamefont {Savard},\ and\ \citenamefont {Thompson}}]{PhysRevLett.100.192502}%
  \BibitemOpen
  \bibfield  {author} {\bibinfo {author} {\bibfnamefont {I.}~\bibnamefont {Tanihata}}, \bibinfo {author} {\bibfnamefont {M.}~\bibnamefont {Alcorta}}, \bibinfo {author} {\bibfnamefont {D.}~\bibnamefont {Bandyopadhyay}}, \bibinfo {author} {\bibfnamefont {R.}~\bibnamefont {Bieri}}, \bibinfo {author} {\bibfnamefont {L.}~\bibnamefont {Buchmann}}, \bibinfo {author} {\bibfnamefont {B.}~\bibnamefont {Davids}}, \bibinfo {author} {\bibfnamefont {N.}~\bibnamefont {Galinski}}, \bibinfo {author} {\bibfnamefont {D.}~\bibnamefont {Howell}}, \bibinfo {author} {\bibfnamefont {W.}~\bibnamefont {Mills}}, \bibinfo {author} {\bibfnamefont {S.}~\bibnamefont {Mythili}}, \bibinfo {author} {\bibfnamefont {R.}~\bibnamefont {Openshaw}}, \bibinfo {author} {\bibfnamefont {E.}~\bibnamefont {Padilla-Rodal}}, \bibinfo {author} {\bibfnamefont {G.}~\bibnamefont {Ruprecht}}, \bibinfo {author} {\bibfnamefont {G.}~\bibnamefont {Sheffer}}, \bibinfo {author} {\bibfnamefont {A.~C.}\ \bibnamefont {Shotter}}, \bibinfo {author} {\bibfnamefont
  {M.}~\bibnamefont {Trinczek}}, \bibinfo {author} {\bibfnamefont {P.}~\bibnamefont {Walden}}, \bibinfo {author} {\bibfnamefont {H.}~\bibnamefont {Savajols}}, \bibinfo {author} {\bibfnamefont {T.}~\bibnamefont {Roger}}, \bibinfo {author} {\bibfnamefont {M.}~\bibnamefont {Caamano}}, \bibinfo {author} {\bibfnamefont {W.}~\bibnamefont {Mittig}}, \bibinfo {author} {\bibfnamefont {P.}~\bibnamefont {Roussel-Chomaz}}, \bibinfo {author} {\bibfnamefont {R.}~\bibnamefont {Kanungo}}, \bibinfo {author} {\bibfnamefont {A.}~\bibnamefont {Gallant}}, \bibinfo {author} {\bibfnamefont {M.}~\bibnamefont {Notani}}, \bibinfo {author} {\bibfnamefont {G.}~\bibnamefont {Savard}}, \ and\ \bibinfo {author} {\bibfnamefont {I.~J.}\ \bibnamefont {Thompson}},\ }\href {\doibase 10.1103/PhysRevLett.100.192502} {\bibfield  {journal} {\bibinfo  {journal} {Phys. Rev. Lett.}\ }\textbf {\bibinfo {volume} {100}},\ \bibinfo {pages} {192502} (\bibinfo {year} {2008})}\BibitemShut {NoStop}%
\bibitem [{\citenamefont {Hamamoto}(2010)}]{PhysRevC.81.021304}%
  \BibitemOpen
  \bibfield  {author} {\bibinfo {author} {\bibfnamefont {I.}~\bibnamefont {Hamamoto}},\ }\href {\doibase 10.1103/PhysRevC.81.021304} {\bibfield  {journal} {\bibinfo  {journal} {Phys. Rev. C}\ }\textbf {\bibinfo {volume} {81}},\ \bibinfo {pages} {021304(R)} (\bibinfo {year} {2010})}\BibitemShut {NoStop}%
\bibitem [{\citenamefont {Tsunoda}\ \emph {et~al.}(2020)\citenamefont {Tsunoda}, \citenamefont {Otsuka}, \citenamefont {Takayanagi}, \citenamefont {Shimizu}, \citenamefont {Suzuki}, \citenamefont {Utsuno}, \citenamefont {Yoshida},\ and\ \citenamefont {Ueno}}]{Tsunoda2020}%
  \BibitemOpen
  \bibfield  {author} {\bibinfo {author} {\bibfnamefont {N.}~\bibnamefont {Tsunoda}}, \bibinfo {author} {\bibfnamefont {T.}~\bibnamefont {Otsuka}}, \bibinfo {author} {\bibfnamefont {K.}~\bibnamefont {Takayanagi}}, \bibinfo {author} {\bibfnamefont {N.}~\bibnamefont {Shimizu}}, \bibinfo {author} {\bibfnamefont {T.}~\bibnamefont {Suzuki}}, \bibinfo {author} {\bibfnamefont {Y.}~\bibnamefont {Utsuno}}, \bibinfo {author} {\bibfnamefont {S.}~\bibnamefont {Yoshida}}, \ and\ \bibinfo {author} {\bibfnamefont {H.}~\bibnamefont {Ueno}},\ }\href {\doibase 10.1038/s41586-020-2848-x} {\bibfield  {journal} {\bibinfo  {journal} {Nature}\ }\textbf {\bibinfo {volume} {587}},\ \bibinfo {pages} {66} (\bibinfo {year} {2020})}\BibitemShut {NoStop}%
\bibitem [{\citenamefont {Zhukov}\ \emph {et~al.}(1993)\citenamefont {Zhukov}, \citenamefont {Danilin}, \citenamefont {Fedorov}, \citenamefont {Bang}, \citenamefont {Thompson},\ and\ \citenamefont {Vaagen}}]{ZHUKOV1993151}%
  \BibitemOpen
  \bibfield  {author} {\bibinfo {author} {\bibfnamefont {M.}~\bibnamefont {Zhukov}}, \bibinfo {author} {\bibfnamefont {B.}~\bibnamefont {Danilin}}, \bibinfo {author} {\bibfnamefont {D.}~\bibnamefont {Fedorov}}, \bibinfo {author} {\bibfnamefont {J.}~\bibnamefont {Bang}}, \bibinfo {author} {\bibfnamefont {I.}~\bibnamefont {Thompson}}, \ and\ \bibinfo {author} {\bibfnamefont {J.}~\bibnamefont {Vaagen}},\ }\href {\doibase https://doi.org/10.1016/0370-1573(93)90141-Y} {\bibfield  {journal} {\bibinfo  {journal} {Physics Reports}\ }\textbf {\bibinfo {volume} {231}},\ \bibinfo {pages} {151} (\bibinfo {year} {1993})}\BibitemShut {NoStop}%
\bibitem [{\citenamefont {Hansen}\ \emph {et~al.}(1995)\citenamefont {Hansen}, \citenamefont {Jensen},\ and\ \citenamefont {Jonson}}]{doi:10.1146/annurev.ns.45.120195.003111}%
  \BibitemOpen
  \bibfield  {author} {\bibinfo {author} {\bibfnamefont {P.~G.}\ \bibnamefont {Hansen}}, \bibinfo {author} {\bibfnamefont {A.~S.}\ \bibnamefont {Jensen}}, \ and\ \bibinfo {author} {\bibfnamefont {B.}~\bibnamefont {Jonson}},\ }\href {\doibase 10.1146/annurev.ns.45.120195.003111} {\bibfield  {journal} {\bibinfo  {journal} {Annual Review of Nuclear and Particle Science}\ }\textbf {\bibinfo {volume} {45}},\ \bibinfo {pages} {591} (\bibinfo {year} {1995})},\ \Eprint {http://arxiv.org/abs/https://doi.org/10.1146/annurev.ns.45.120195.003111} {https://doi.org/10.1146/annurev.ns.45.120195.003111} \BibitemShut {NoStop}%
\bibitem [{\citenamefont {Otsuka}\ \emph {et~al.}(1993)\citenamefont {Otsuka}, \citenamefont {Fukunishi},\ and\ \citenamefont {Sagawa}}]{PhysRevLett.70.1385}%
  \BibitemOpen
  \bibfield  {author} {\bibinfo {author} {\bibfnamefont {T.}~\bibnamefont {Otsuka}}, \bibinfo {author} {\bibfnamefont {N.}~\bibnamefont {Fukunishi}}, \ and\ \bibinfo {author} {\bibfnamefont {H.}~\bibnamefont {Sagawa}},\ }\href {\doibase 10.1103/PhysRevLett.70.1385} {\bibfield  {journal} {\bibinfo  {journal} {Phys. Rev. Lett.}\ }\textbf {\bibinfo {volume} {70}},\ \bibinfo {pages} {1385} (\bibinfo {year} {1993})}\BibitemShut {NoStop}%
\bibitem [{\citenamefont {Kuo}\ \emph {et~al.}(1997)\citenamefont {Kuo}, \citenamefont {Krmpoti\ifmmode~\acute{c}\else \'{c}\fi{}},\ and\ \citenamefont {Tzeng}}]{PhysRevLett.78.2708}%
  \BibitemOpen
  \bibfield  {author} {\bibinfo {author} {\bibfnamefont {T.~T.~S.}\ \bibnamefont {Kuo}}, \bibinfo {author} {\bibfnamefont {F.}~\bibnamefont {Krmpoti\ifmmode~\acute{c}\else \'{c}\fi{}}}, \ and\ \bibinfo {author} {\bibfnamefont {Y.}~\bibnamefont {Tzeng}},\ }\href {\doibase 10.1103/PhysRevLett.78.2708} {\bibfield  {journal} {\bibinfo  {journal} {Phys. Rev. Lett.}\ }\textbf {\bibinfo {volume} {78}},\ \bibinfo {pages} {2708} (\bibinfo {year} {1997})}\BibitemShut {NoStop}%
\bibitem [{\citenamefont {Itagaki}\ and\ \citenamefont {Aoyama}(1999)}]{PhysRevC.61.024303}%
  \BibitemOpen
  \bibfield  {author} {\bibinfo {author} {\bibfnamefont {N.}~\bibnamefont {Itagaki}}\ and\ \bibinfo {author} {\bibfnamefont {S.}~\bibnamefont {Aoyama}},\ }\href {\doibase 10.1103/PhysRevC.61.024303} {\bibfield  {journal} {\bibinfo  {journal} {Phys. Rev. C}\ }\textbf {\bibinfo {volume} {61}},\ \bibinfo {pages} {024303} (\bibinfo {year} {1999})}\BibitemShut {NoStop}%
\bibitem [{\citenamefont {Ryberg}\ \emph {et~al.}(2014)\citenamefont {Ryberg}, \citenamefont {Forss\'en}, \citenamefont {Hammer},\ and\ \citenamefont {Platter}}]{PhysRevC.89.014325}%
  \BibitemOpen
  \bibfield  {author} {\bibinfo {author} {\bibfnamefont {E.}~\bibnamefont {Ryberg}}, \bibinfo {author} {\bibfnamefont {C.}~\bibnamefont {Forss\'en}}, \bibinfo {author} {\bibfnamefont {H.-W.}\ \bibnamefont {Hammer}}, \ and\ \bibinfo {author} {\bibfnamefont {L.}~\bibnamefont {Platter}},\ }\href {\doibase 10.1103/PhysRevC.89.014325} {\bibfield  {journal} {\bibinfo  {journal} {Phys. Rev. C}\ }\textbf {\bibinfo {volume} {89}},\ \bibinfo {pages} {014325} (\bibinfo {year} {2014})}\BibitemShut {NoStop}%
\bibitem [{\citenamefont {Terasaki}\ \emph {et~al.}(1996)\citenamefont {Terasaki}, \citenamefont {Heenen}, \citenamefont {Flocard},\ and\ \citenamefont {Bonche}}]{TERASAKI1996371}%
  \BibitemOpen
  \bibfield  {author} {\bibinfo {author} {\bibfnamefont {J.}~\bibnamefont {Terasaki}}, \bibinfo {author} {\bibfnamefont {P.-H.}\ \bibnamefont {Heenen}}, \bibinfo {author} {\bibfnamefont {H.}~\bibnamefont {Flocard}}, \ and\ \bibinfo {author} {\bibfnamefont {P.}~\bibnamefont {Bonche}},\ }\href {\doibase https://doi.org/10.1016/0375-9474(96)00036-X} {\bibfield  {journal} {\bibinfo  {journal} {Nuclear Physics A}\ }\textbf {\bibinfo {volume} {600}},\ \bibinfo {pages} {371} (\bibinfo {year} {1996})}\BibitemShut {NoStop}%
\bibitem [{\citenamefont {Schunck}\ and\ \citenamefont {Egido}(2008)}]{PhysRevC.78.064305}%
  \BibitemOpen
  \bibfield  {author} {\bibinfo {author} {\bibfnamefont {N.}~\bibnamefont {Schunck}}\ and\ \bibinfo {author} {\bibfnamefont {J.~L.}\ \bibnamefont {Egido}},\ }\href {\doibase 10.1103/PhysRevC.78.064305} {\bibfield  {journal} {\bibinfo  {journal} {Phys. Rev. C}\ }\textbf {\bibinfo {volume} {78}},\ \bibinfo {pages} {064305} (\bibinfo {year} {2008})}\BibitemShut {NoStop}%
\bibitem [{\citenamefont {Pei}\ \emph {et~al.}(2011)\citenamefont {Pei}, \citenamefont {Kruppa},\ and\ \citenamefont {Nazarewicz}}]{PhysRevC.84.024311}%
  \BibitemOpen
  \bibfield  {author} {\bibinfo {author} {\bibfnamefont {J.~C.}\ \bibnamefont {Pei}}, \bibinfo {author} {\bibfnamefont {A.~T.}\ \bibnamefont {Kruppa}}, \ and\ \bibinfo {author} {\bibfnamefont {W.}~\bibnamefont {Nazarewicz}},\ }\href {\doibase 10.1103/PhysRevC.84.024311} {\bibfield  {journal} {\bibinfo  {journal} {Phys. Rev. C}\ }\textbf {\bibinfo {volume} {84}},\ \bibinfo {pages} {024311} (\bibinfo {year} {2011})}\BibitemShut {NoStop}%
\bibitem [{\citenamefont {Zhang}\ \emph {et~al.}(2013)\citenamefont {Zhang}, \citenamefont {Pei},\ and\ \citenamefont {Xu}}]{PhysRevC.88.054305}%
  \BibitemOpen
  \bibfield  {author} {\bibinfo {author} {\bibfnamefont {Y.~N.}\ \bibnamefont {Zhang}}, \bibinfo {author} {\bibfnamefont {J.~C.}\ \bibnamefont {Pei}}, \ and\ \bibinfo {author} {\bibfnamefont {F.~R.}\ \bibnamefont {Xu}},\ }\href {\doibase 10.1103/PhysRevC.88.054305} {\bibfield  {journal} {\bibinfo  {journal} {Phys. Rev. C}\ }\textbf {\bibinfo {volume} {88}},\ \bibinfo {pages} {054305} (\bibinfo {year} {2013})}\BibitemShut {NoStop}%
\bibitem [{\citenamefont {Tsukiyama}\ \emph {et~al.}(2012)\citenamefont {Tsukiyama}, \citenamefont {Bogner},\ and\ \citenamefont {Schwenk}}]{PhysRevC.85.061304}%
  \BibitemOpen
  \bibfield  {author} {\bibinfo {author} {\bibfnamefont {K.}~\bibnamefont {Tsukiyama}}, \bibinfo {author} {\bibfnamefont {S.~K.}\ \bibnamefont {Bogner}}, \ and\ \bibinfo {author} {\bibfnamefont {A.}~\bibnamefont {Schwenk}},\ }\href {\doibase 10.1103/PhysRevC.85.061304} {\bibfield  {journal} {\bibinfo  {journal} {Phys. Rev. C}\ }\textbf {\bibinfo {volume} {85}},\ \bibinfo {pages} {061304(R)} (\bibinfo {year} {2012})}\BibitemShut {NoStop}%
\bibitem [{\citenamefont {Hergert}\ \emph {et~al.}(2016)\citenamefont {Hergert}, \citenamefont {Bogner}, \citenamefont {Morris}, \citenamefont {Schwenk},\ and\ \citenamefont {Tsukiyama}}]{HERGERT2016165}%
  \BibitemOpen
  \bibfield  {author} {\bibinfo {author} {\bibfnamefont {H.}~\bibnamefont {Hergert}}, \bibinfo {author} {\bibfnamefont {S.}~\bibnamefont {Bogner}}, \bibinfo {author} {\bibfnamefont {T.}~\bibnamefont {Morris}}, \bibinfo {author} {\bibfnamefont {A.}~\bibnamefont {Schwenk}}, \ and\ \bibinfo {author} {\bibfnamefont {K.}~\bibnamefont {Tsukiyama}},\ }\href {\doibase https://doi.org/10.1016/j.physrep.2015.12.007} {\bibfield  {journal} {\bibinfo  {journal} {Physics Reports}\ }\textbf {\bibinfo {volume} {621}},\ \bibinfo {pages} {165} (\bibinfo {year} {2016})},\ \bibinfo {note} {memorial Volume in Honor of Gerald E. Brown}\BibitemShut {NoStop}%
\bibitem [{\citenamefont {Stroberg}\ \emph {et~al.}(2017)\citenamefont {Stroberg}, \citenamefont {Calci}, \citenamefont {Hergert}, \citenamefont {Holt}, \citenamefont {Bogner}, \citenamefont {Roth},\ and\ \citenamefont {Schwenk}}]{PhysRevLett.118.032502}%
  \BibitemOpen
  \bibfield  {author} {\bibinfo {author} {\bibfnamefont {S.~R.}\ \bibnamefont {Stroberg}}, \bibinfo {author} {\bibfnamefont {A.}~\bibnamefont {Calci}}, \bibinfo {author} {\bibfnamefont {H.}~\bibnamefont {Hergert}}, \bibinfo {author} {\bibfnamefont {J.~D.}\ \bibnamefont {Holt}}, \bibinfo {author} {\bibfnamefont {S.~K.}\ \bibnamefont {Bogner}}, \bibinfo {author} {\bibfnamefont {R.}~\bibnamefont {Roth}}, \ and\ \bibinfo {author} {\bibfnamefont {A.}~\bibnamefont {Schwenk}},\ }\href {\doibase 10.1103/PhysRevLett.118.032502} {\bibfield  {journal} {\bibinfo  {journal} {Phys. Rev. Lett.}\ }\textbf {\bibinfo {volume} {118}},\ \bibinfo {pages} {032502} (\bibinfo {year} {2017})}\BibitemShut {NoStop}%
\bibitem [{\citenamefont {Li}\ \emph {et~al.}(2023)\citenamefont {Li}, \citenamefont {Yuan}, \citenamefont {Li}, \citenamefont {Xie}, \citenamefont {Zhang}, \citenamefont {Zhang}, \citenamefont {Xu}, \citenamefont {Michel}, \citenamefont {Xu},\ and\ \citenamefont {Zuo}}]{PhysRevC.107.014302}%
  \BibitemOpen
  \bibfield  {author} {\bibinfo {author} {\bibfnamefont {H.~H.}\ \bibnamefont {Li}}, \bibinfo {author} {\bibfnamefont {Q.}~\bibnamefont {Yuan}}, \bibinfo {author} {\bibfnamefont {J.~G.}\ \bibnamefont {Li}}, \bibinfo {author} {\bibfnamefont {M.~R.}\ \bibnamefont {Xie}}, \bibinfo {author} {\bibfnamefont {S.}~\bibnamefont {Zhang}}, \bibinfo {author} {\bibfnamefont {Y.~H.}\ \bibnamefont {Zhang}}, \bibinfo {author} {\bibfnamefont {X.~X.}\ \bibnamefont {Xu}}, \bibinfo {author} {\bibfnamefont {N.}~\bibnamefont {Michel}}, \bibinfo {author} {\bibfnamefont {F.~R.}\ \bibnamefont {Xu}}, \ and\ \bibinfo {author} {\bibfnamefont {W.}~\bibnamefont {Zuo}},\ }\href {\doibase 10.1103/PhysRevC.107.014302} {\bibfield  {journal} {\bibinfo  {journal} {Phys. Rev. C}\ }\textbf {\bibinfo {volume} {107}},\ \bibinfo {pages} {014302} (\bibinfo {year} {2023})}\BibitemShut {NoStop}%
\bibitem [{\citenamefont {Stroberg}\ \emph {et~al.}(2019)\citenamefont {Stroberg}, \citenamefont {Hergert}, \citenamefont {Bogner},\ and\ \citenamefont {Holt}}]{doi:10.1146/annurev-nucl-101917-021120}%
  \BibitemOpen
  \bibfield  {author} {\bibinfo {author} {\bibfnamefont {S.~R.}\ \bibnamefont {Stroberg}}, \bibinfo {author} {\bibfnamefont {H.}~\bibnamefont {Hergert}}, \bibinfo {author} {\bibfnamefont {S.~K.}\ \bibnamefont {Bogner}}, \ and\ \bibinfo {author} {\bibfnamefont {J.~D.}\ \bibnamefont {Holt}},\ }\href {\doibase 10.1146/annurev-nucl-101917-021120} {\bibfield  {journal} {\bibinfo  {journal} {Annual Review of Nuclear and Particle Science}\ }\textbf {\bibinfo {volume} {69}},\ \bibinfo {pages} {307} (\bibinfo {year} {2019})},\ \Eprint {http://arxiv.org/abs/https://doi.org/10.1146/annurev-nucl-101917-021120} {https://doi.org/10.1146/annurev-nucl-101917-021120} \BibitemShut {NoStop}%
\bibitem [{\citenamefont {Wegner}(2001)}]{WEGNER200177}%
  \BibitemOpen
  \bibfield  {author} {\bibinfo {author} {\bibfnamefont {F.~J.}\ \bibnamefont {Wegner}},\ }\href {\doibase https://doi.org/10.1016/S0370-1573(00)00136-8} {\bibfield  {journal} {\bibinfo  {journal} {Physics Reports}\ }\textbf {\bibinfo {volume} {348}},\ \bibinfo {pages} {77} (\bibinfo {year} {2001})},\ \bibinfo {note} {renormalization group theory in the new millennium. II}\BibitemShut {NoStop}%
\bibitem [{\citenamefont {Li}\ \emph {et~al.}(2022)\citenamefont {Li}, \citenamefont {Michel}, \citenamefont {Li},\ and\ \citenamefont {Zuo}}]{LI2022137225}%
  \BibitemOpen
  \bibfield  {author} {\bibinfo {author} {\bibfnamefont {J.}~\bibnamefont {Li}}, \bibinfo {author} {\bibfnamefont {N.}~\bibnamefont {Michel}}, \bibinfo {author} {\bibfnamefont {H.}~\bibnamefont {Li}}, \ and\ \bibinfo {author} {\bibfnamefont {W.}~\bibnamefont {Zuo}},\ }\href {\doibase https://doi.org/10.1016/j.physletb.2022.137225} {\bibfield  {journal} {\bibinfo  {journal} {Physics Letters B}\ }\textbf {\bibinfo {volume} {832}},\ \bibinfo {pages} {137225} (\bibinfo {year} {2022})}\BibitemShut {NoStop}%
\bibitem [{\citenamefont {Li}\ \emph {et~al.}(2012)\citenamefont {Li}, \citenamefont {Meng}, \citenamefont {Ring}, \citenamefont {Zhao},\ and\ \citenamefont {Zhou}}]{PhysRevC.85.024312}%
  \BibitemOpen
  \bibfield  {author} {\bibinfo {author} {\bibfnamefont {L.}~\bibnamefont {Li}}, \bibinfo {author} {\bibfnamefont {J.}~\bibnamefont {Meng}}, \bibinfo {author} {\bibfnamefont {P.}~\bibnamefont {Ring}}, \bibinfo {author} {\bibfnamefont {E.-G.}\ \bibnamefont {Zhao}}, \ and\ \bibinfo {author} {\bibfnamefont {S.-G.}\ \bibnamefont {Zhou}},\ }\href {\doibase 10.1103/PhysRevC.85.024312} {\bibfield  {journal} {\bibinfo  {journal} {Phys. Rev. C}\ }\textbf {\bibinfo {volume} {85}},\ \bibinfo {pages} {024312} (\bibinfo {year} {2012})}\BibitemShut {NoStop}%
\bibitem [{\citenamefont {Michel}\ \emph {et~al.}(2020)\citenamefont {Michel}, \citenamefont {Li}, \citenamefont {Xu},\ and\ \citenamefont {Zuo}}]{PhysRevC.101.031301}%
  \BibitemOpen
  \bibfield  {author} {\bibinfo {author} {\bibfnamefont {N.}~\bibnamefont {Michel}}, \bibinfo {author} {\bibfnamefont {J.~G.}\ \bibnamefont {Li}}, \bibinfo {author} {\bibfnamefont {F.~R.}\ \bibnamefont {Xu}}, \ and\ \bibinfo {author} {\bibfnamefont {W.}~\bibnamefont {Zuo}},\ }\href {\doibase 10.1103/PhysRevC.101.031301} {\bibfield  {journal} {\bibinfo  {journal} {Phys. Rev. C}\ }\textbf {\bibinfo {volume} {101}},\ \bibinfo {pages} {031301(R)} (\bibinfo {year} {2020})}\BibitemShut {NoStop}%
\bibitem [{\citenamefont {Zhou}\ \emph {et~al.}(2010)\citenamefont {Zhou}, \citenamefont {Meng}, \citenamefont {Ring},\ and\ \citenamefont {Zhao}}]{PhysRevC.82.011301}%
  \BibitemOpen
  \bibfield  {author} {\bibinfo {author} {\bibfnamefont {S.-G.}\ \bibnamefont {Zhou}}, \bibinfo {author} {\bibfnamefont {J.}~\bibnamefont {Meng}}, \bibinfo {author} {\bibfnamefont {P.}~\bibnamefont {Ring}}, \ and\ \bibinfo {author} {\bibfnamefont {E.-G.}\ \bibnamefont {Zhao}},\ }\href {\doibase 10.1103/PhysRevC.82.011301} {\bibfield  {journal} {\bibinfo  {journal} {Phys. Rev. C}\ }\textbf {\bibinfo {volume} {82}},\ \bibinfo {pages} {011301(R)} (\bibinfo {year} {2010})}\BibitemShut {NoStop}%
\bibitem [{\citenamefont {Sun}\ and\ \citenamefont {Zhou}(2021)}]{SUN20212072}%
  \BibitemOpen
  \bibfield  {author} {\bibinfo {author} {\bibfnamefont {X.-X.}\ \bibnamefont {Sun}}\ and\ \bibinfo {author} {\bibfnamefont {S.-G.}\ \bibnamefont {Zhou}},\ }\href {\doibase https://doi.org/10.1016/j.scib.2021.07.005} {\bibfield  {journal} {\bibinfo  {journal} {Science Bulletin}\ }\textbf {\bibinfo {volume} {66}},\ \bibinfo {pages} {2072} (\bibinfo {year} {2021})}\BibitemShut {NoStop}%
\bibitem [{\citenamefont {Simpson}\ and\ \citenamefont {Tostevin}(2010)}]{PhysRevC.82.044616}%
  \BibitemOpen
  \bibfield  {author} {\bibinfo {author} {\bibfnamefont {E.~C.}\ \bibnamefont {Simpson}}\ and\ \bibinfo {author} {\bibfnamefont {J.~A.}\ \bibnamefont {Tostevin}},\ }\href {\doibase 10.1103/PhysRevC.82.044616} {\bibfield  {journal} {\bibinfo  {journal} {Phys. Rev. C}\ }\textbf {\bibinfo {volume} {82}},\ \bibinfo {pages} {044616} (\bibinfo {year} {2010})}\BibitemShut {NoStop}%
\bibitem [{\citenamefont {Hebeler}\ \emph {et~al.}(2011)\citenamefont {Hebeler}, \citenamefont {Bogner}, \citenamefont {Furnstahl}, \citenamefont {Nogga},\ and\ \citenamefont {Schwenk}}]{PhysRevC.83.031301}%
  \BibitemOpen
  \bibfield  {author} {\bibinfo {author} {\bibfnamefont {K.}~\bibnamefont {Hebeler}}, \bibinfo {author} {\bibfnamefont {S.~K.}\ \bibnamefont {Bogner}}, \bibinfo {author} {\bibfnamefont {R.~J.}\ \bibnamefont {Furnstahl}}, \bibinfo {author} {\bibfnamefont {A.}~\bibnamefont {Nogga}}, \ and\ \bibinfo {author} {\bibfnamefont {A.}~\bibnamefont {Schwenk}},\ }\href {\doibase 10.1103/PhysRevC.83.031301} {\bibfield  {journal} {\bibinfo  {journal} {Phys. Rev. C}\ }\textbf {\bibinfo {volume} {83}},\ \bibinfo {pages} {031301(R)} (\bibinfo {year} {2011})}\BibitemShut {NoStop}%
\bibitem [{\citenamefont {Simonis}\ \emph {et~al.}(2016)\citenamefont {Simonis}, \citenamefont {Hebeler}, \citenamefont {Holt}, \citenamefont {Men\'endez},\ and\ \citenamefont {Schwenk}}]{PhysRevC.93.011302}%
  \BibitemOpen
  \bibfield  {author} {\bibinfo {author} {\bibfnamefont {J.}~\bibnamefont {Simonis}}, \bibinfo {author} {\bibfnamefont {K.}~\bibnamefont {Hebeler}}, \bibinfo {author} {\bibfnamefont {J.~D.}\ \bibnamefont {Holt}}, \bibinfo {author} {\bibfnamefont {J.}~\bibnamefont {Men\'endez}}, \ and\ \bibinfo {author} {\bibfnamefont {A.}~\bibnamefont {Schwenk}},\ }\href {\doibase 10.1103/PhysRevC.93.011302} {\bibfield  {journal} {\bibinfo  {journal} {Phys. Rev. C}\ }\textbf {\bibinfo {volume} {93}},\ \bibinfo {pages} {011302(R)} (\bibinfo {year} {2016})}\BibitemShut {NoStop}%
\bibitem [{\citenamefont {Som\`a}\ \emph {et~al.}(2020)\citenamefont {Som\`a}, \citenamefont {Navr\'atil}, \citenamefont {Raimondi}, \citenamefont {Barbieri},\ and\ \citenamefont {Duguet}}]{PhysRevC.101.014318}%
  \BibitemOpen
  \bibfield  {author} {\bibinfo {author} {\bibfnamefont {V.}~\bibnamefont {Som\`a}}, \bibinfo {author} {\bibfnamefont {P.}~\bibnamefont {Navr\'atil}}, \bibinfo {author} {\bibfnamefont {F.}~\bibnamefont {Raimondi}}, \bibinfo {author} {\bibfnamefont {C.}~\bibnamefont {Barbieri}}, \ and\ \bibinfo {author} {\bibfnamefont {T.}~\bibnamefont {Duguet}},\ }\href {\doibase 10.1103/PhysRevC.101.014318} {\bibfield  {journal} {\bibinfo  {journal} {Phys. Rev. C}\ }\textbf {\bibinfo {volume} {101}},\ \bibinfo {pages} {014318} (\bibinfo {year} {2020})}\BibitemShut {NoStop}%
\bibitem [{\citenamefont {Yuan}\ \emph {et~al.}(2024)\citenamefont {Yuan}, \citenamefont {Li},\ and\ \citenamefont {Li}}]{YUAN2024138331}%
  \BibitemOpen
  \bibfield  {author} {\bibinfo {author} {\bibfnamefont {Q.}~\bibnamefont {Yuan}}, \bibinfo {author} {\bibfnamefont {J.}~\bibnamefont {Li}}, \ and\ \bibinfo {author} {\bibfnamefont {H.}~\bibnamefont {Li}},\ }\href {\doibase https://doi.org/10.1016/j.physletb.2023.138331} {\bibfield  {journal} {\bibinfo  {journal} {Physics Letters B}\ }\textbf {\bibinfo {volume} {848}},\ \bibinfo {pages} {138331} (\bibinfo {year} {2024})}\BibitemShut {NoStop}%
\bibitem [{\citenamefont {Roth}\ \emph {et~al.}(2012)\citenamefont {Roth}, \citenamefont {Binder}, \citenamefont {Vobig}, \citenamefont {Calci}, \citenamefont {Langhammer},\ and\ \citenamefont {Navr\'atil}}]{PhysRevLett.109.052501}%
  \BibitemOpen
  \bibfield  {author} {\bibinfo {author} {\bibfnamefont {R.}~\bibnamefont {Roth}}, \bibinfo {author} {\bibfnamefont {S.}~\bibnamefont {Binder}}, \bibinfo {author} {\bibfnamefont {K.}~\bibnamefont {Vobig}}, \bibinfo {author} {\bibfnamefont {A.}~\bibnamefont {Calci}}, \bibinfo {author} {\bibfnamefont {J.}~\bibnamefont {Langhammer}}, \ and\ \bibinfo {author} {\bibfnamefont {P.}~\bibnamefont {Navr\'atil}},\ }\href {\doibase 10.1103/PhysRevLett.109.052501} {\bibfield  {journal} {\bibinfo  {journal} {Phys. Rev. Lett.}\ }\textbf {\bibinfo {volume} {109}},\ \bibinfo {pages} {052501} (\bibinfo {year} {2012})}\BibitemShut {NoStop}%
\bibitem [{\citenamefont {Morris}\ \emph {et~al.}(2015)\citenamefont {Morris}, \citenamefont {Parzuchowski},\ and\ \citenamefont {Bogner}}]{PhysRevC.92.034331}%
  \BibitemOpen
  \bibfield  {author} {\bibinfo {author} {\bibfnamefont {T.~D.}\ \bibnamefont {Morris}}, \bibinfo {author} {\bibfnamefont {N.~M.}\ \bibnamefont {Parzuchowski}}, \ and\ \bibinfo {author} {\bibfnamefont {S.~K.}\ \bibnamefont {Bogner}},\ }\href {\doibase 10.1103/PhysRevC.92.034331} {\bibfield  {journal} {\bibinfo  {journal} {Phys. Rev. C}\ }\textbf {\bibinfo {volume} {92}},\ \bibinfo {pages} {034331} (\bibinfo {year} {2015})}\BibitemShut {NoStop}%
\bibitem [{\citenamefont {Shimizu}\ \emph {et~al.}(2019)\citenamefont {Shimizu}, \citenamefont {Mizusaki}, \citenamefont {Utsuno},\ and\ \citenamefont {Tsunoda}}]{SHIMIZU2019372}%
  \BibitemOpen
  \bibfield  {author} {\bibinfo {author} {\bibfnamefont {N.}~\bibnamefont {Shimizu}}, \bibinfo {author} {\bibfnamefont {T.}~\bibnamefont {Mizusaki}}, \bibinfo {author} {\bibfnamefont {Y.}~\bibnamefont {Utsuno}}, \ and\ \bibinfo {author} {\bibfnamefont {Y.}~\bibnamefont {Tsunoda}},\ }\href {\doibase https://doi.org/10.1016/j.cpc.2019.06.011} {\bibfield  {journal} {\bibinfo  {journal} {Computer Physics Communications}\ }\textbf {\bibinfo {volume} {244}},\ \bibinfo {pages} {372} (\bibinfo {year} {2019})}\BibitemShut {NoStop}%
\bibitem [{\citenamefont {Wang}\ \emph {et~al.}(2021)\citenamefont {Wang}, \citenamefont {Huang}, \citenamefont {Kondev}, \citenamefont {Audi},\ and\ \citenamefont {Naimi}}]{Wang_2021}%
  \BibitemOpen
  \bibfield  {author} {\bibinfo {author} {\bibfnamefont {M.}~\bibnamefont {Wang}}, \bibinfo {author} {\bibfnamefont {W.}~\bibnamefont {Huang}}, \bibinfo {author} {\bibfnamefont {F.}~\bibnamefont {Kondev}}, \bibinfo {author} {\bibfnamefont {G.}~\bibnamefont {Audi}}, \ and\ \bibinfo {author} {\bibfnamefont {S.}~\bibnamefont {Naimi}},\ }\href {\doibase 10.1088/1674-1137/abddaf} {\bibfield  {journal} {\bibinfo  {journal} {Chinese Physics C}\ }\textbf {\bibinfo {volume} {45}},\ \bibinfo {pages} {030003} (\bibinfo {year} {2021})}\BibitemShut {NoStop}%
\bibitem [{ens()}]{ensdf}%
  \BibitemOpen
  \href@noop {} {}\bibinfo {howpublished} {\url{http://www.nndc.bnl.gov/ensdf}}\BibitemShut {NoStop}%
\bibitem [{\citenamefont {Möller}\ \emph {et~al.}(2016)\citenamefont {Möller}, \citenamefont {Sierk}, \citenamefont {Ichikawa},\ and\ \citenamefont {Sagawa}}]{MOLLER20161}%
  \BibitemOpen
  \bibfield  {author} {\bibinfo {author} {\bibfnamefont {P.}~\bibnamefont {Möller}}, \bibinfo {author} {\bibfnamefont {A.}~\bibnamefont {Sierk}}, \bibinfo {author} {\bibfnamefont {T.}~\bibnamefont {Ichikawa}}, \ and\ \bibinfo {author} {\bibfnamefont {H.}~\bibnamefont {Sagawa}},\ }\href {\doibase https://doi.org/10.1016/j.adt.2015.10.002} {\bibfield  {journal} {\bibinfo  {journal} {Atomic Data and Nuclear Data Tables}\ }\textbf {\bibinfo {volume} {109-110}},\ \bibinfo {pages} {1} (\bibinfo {year} {2016})}\BibitemShut {NoStop}%
\bibitem [{HFB()}]{HFBgogny}%
  \BibitemOpen
  \href@noop {} {}\bibinfo {howpublished} {\url{https://www-phynu.cea.fr/science_en_ligne/carte_potentiels_microscopiques/noyaux/zz12/sep/zz12nn27sep_eng.html}}\BibitemShut {NoStop}%
\bibitem [{DFT()}]{DFTSkyrme}%
  \BibitemOpen
  \href@noop {} {}\bibinfo {howpublished} {\url{https://massexplorer.frib.msu.edu/content/DFTMassTables.html}}\BibitemShut {NoStop}%
\bibitem [{\citenamefont {Suzuki}\ \emph {et~al.}(2002)\citenamefont {Suzuki}, \citenamefont {Ogawa}, \citenamefont {Chiba}, \citenamefont {Fukuda}, \citenamefont {Iwasa}, \citenamefont {Izumikawa}, \citenamefont {Kanungo}, \citenamefont {Kawamura}, \citenamefont {Ozawa}, \citenamefont {Suda}, \citenamefont {Tanihata}, \citenamefont {Watanabe}, \citenamefont {Yamaguchi},\ and\ \citenamefont {Yamaguchi}}]{PhysRevLett.89.012501}%
  \BibitemOpen
  \bibfield  {author} {\bibinfo {author} {\bibfnamefont {T.}~\bibnamefont {Suzuki}}, \bibinfo {author} {\bibfnamefont {Y.}~\bibnamefont {Ogawa}}, \bibinfo {author} {\bibfnamefont {M.}~\bibnamefont {Chiba}}, \bibinfo {author} {\bibfnamefont {M.}~\bibnamefont {Fukuda}}, \bibinfo {author} {\bibfnamefont {N.}~\bibnamefont {Iwasa}}, \bibinfo {author} {\bibfnamefont {T.}~\bibnamefont {Izumikawa}}, \bibinfo {author} {\bibfnamefont {R.}~\bibnamefont {Kanungo}}, \bibinfo {author} {\bibfnamefont {Y.}~\bibnamefont {Kawamura}}, \bibinfo {author} {\bibfnamefont {A.}~\bibnamefont {Ozawa}}, \bibinfo {author} {\bibfnamefont {T.}~\bibnamefont {Suda}}, \bibinfo {author} {\bibfnamefont {I.}~\bibnamefont {Tanihata}}, \bibinfo {author} {\bibfnamefont {S.}~\bibnamefont {Watanabe}}, \bibinfo {author} {\bibfnamefont {T.}~\bibnamefont {Yamaguchi}}, \ and\ \bibinfo {author} {\bibfnamefont {Y.}~\bibnamefont {Yamaguchi}},\ }\href {\doibase 10.1103/PhysRevLett.89.012501} {\bibfield  {journal} {\bibinfo  {journal} {Phys. Rev. Lett.}\
  }\textbf {\bibinfo {volume} {89}},\ \bibinfo {pages} {012501} (\bibinfo {year} {2002})}\BibitemShut {NoStop}%
\bibitem [{\citenamefont {Gaudefroy}\ \emph {et~al.}(2012)\citenamefont {Gaudefroy}, \citenamefont {Mittig}, \citenamefont {Orr}, \citenamefont {Varet}, \citenamefont {Chartier}, \citenamefont {Roussel-Chomaz}, \citenamefont {Ebran}, \citenamefont {Fern\'andez-Dom\'{\i}nguez}, \citenamefont {Fr\'emont}, \citenamefont {Gangnant}, \citenamefont {Gillibert}, \citenamefont {Gr\'evy}, \citenamefont {Libin}, \citenamefont {Maslov}, \citenamefont {Paschalis}, \citenamefont {Pietras}, \citenamefont {Penionzhkevich}, \citenamefont {Spitaels},\ and\ \citenamefont {Villari}}]{PhysRevLett.109.202503}%
  \BibitemOpen
  \bibfield  {author} {\bibinfo {author} {\bibfnamefont {L.}~\bibnamefont {Gaudefroy}}, \bibinfo {author} {\bibfnamefont {W.}~\bibnamefont {Mittig}}, \bibinfo {author} {\bibfnamefont {N.~A.}\ \bibnamefont {Orr}}, \bibinfo {author} {\bibfnamefont {S.}~\bibnamefont {Varet}}, \bibinfo {author} {\bibfnamefont {M.}~\bibnamefont {Chartier}}, \bibinfo {author} {\bibfnamefont {P.}~\bibnamefont {Roussel-Chomaz}}, \bibinfo {author} {\bibfnamefont {J.~P.}\ \bibnamefont {Ebran}}, \bibinfo {author} {\bibfnamefont {B.}~\bibnamefont {Fern\'andez-Dom\'{\i}nguez}}, \bibinfo {author} {\bibfnamefont {G.}~\bibnamefont {Fr\'emont}}, \bibinfo {author} {\bibfnamefont {P.}~\bibnamefont {Gangnant}}, \bibinfo {author} {\bibfnamefont {A.}~\bibnamefont {Gillibert}}, \bibinfo {author} {\bibfnamefont {S.}~\bibnamefont {Gr\'evy}}, \bibinfo {author} {\bibfnamefont {J.~F.}\ \bibnamefont {Libin}}, \bibinfo {author} {\bibfnamefont {V.~A.}\ \bibnamefont {Maslov}}, \bibinfo {author} {\bibfnamefont {S.}~\bibnamefont {Paschalis}}, \bibinfo
  {author} {\bibfnamefont {B.}~\bibnamefont {Pietras}}, \bibinfo {author} {\bibfnamefont {Y.-E.}\ \bibnamefont {Penionzhkevich}}, \bibinfo {author} {\bibfnamefont {C.}~\bibnamefont {Spitaels}}, \ and\ \bibinfo {author} {\bibfnamefont {A.~C.~C.}\ \bibnamefont {Villari}},\ }\href {\doibase 10.1103/PhysRevLett.109.202503} {\bibfield  {journal} {\bibinfo  {journal} {Phys. Rev. Lett.}\ }\textbf {\bibinfo {volume} {109}},\ \bibinfo {pages} {202503} (\bibinfo {year} {2012})}\BibitemShut {NoStop}%
\bibitem [{\citenamefont {Kobayashi}\ \emph {et~al.}(2016)\citenamefont {Kobayashi}, \citenamefont {Nakamura}, \citenamefont {Kondo}, \citenamefont {Tostevin}, \citenamefont {Aoi}, \citenamefont {Baba}, \citenamefont {Barthelemy}, \citenamefont {Famiano}, \citenamefont {Fukuda}, \citenamefont {Inabe}, \citenamefont {Ishihara}, \citenamefont {Kanungo}, \citenamefont {Kim}, \citenamefont {Kubo}, \citenamefont {Lee}, \citenamefont {Lee}, \citenamefont {Matsushita}, \citenamefont {Motobayashi}, \citenamefont {Ohnishi}, \citenamefont {Orr}, \citenamefont {Otsu}, \citenamefont {Sako}, \citenamefont {Sakurai}, \citenamefont {Satou}, \citenamefont {Sumikama}, \citenamefont {Takeda}, \citenamefont {Takeuchi}, \citenamefont {Tanaka}, \citenamefont {Togano},\ and\ \citenamefont {Yoneda}}]{PhysRevC.93.014613}%
  \BibitemOpen
  \bibfield  {author} {\bibinfo {author} {\bibfnamefont {N.}~\bibnamefont {Kobayashi}}, \bibinfo {author} {\bibfnamefont {T.}~\bibnamefont {Nakamura}}, \bibinfo {author} {\bibfnamefont {Y.}~\bibnamefont {Kondo}}, \bibinfo {author} {\bibfnamefont {J.~A.}\ \bibnamefont {Tostevin}}, \bibinfo {author} {\bibfnamefont {N.}~\bibnamefont {Aoi}}, \bibinfo {author} {\bibfnamefont {H.}~\bibnamefont {Baba}}, \bibinfo {author} {\bibfnamefont {R.}~\bibnamefont {Barthelemy}}, \bibinfo {author} {\bibfnamefont {M.~A.}\ \bibnamefont {Famiano}}, \bibinfo {author} {\bibfnamefont {N.}~\bibnamefont {Fukuda}}, \bibinfo {author} {\bibfnamefont {N.}~\bibnamefont {Inabe}}, \bibinfo {author} {\bibfnamefont {M.}~\bibnamefont {Ishihara}}, \bibinfo {author} {\bibfnamefont {R.}~\bibnamefont {Kanungo}}, \bibinfo {author} {\bibfnamefont {S.}~\bibnamefont {Kim}}, \bibinfo {author} {\bibfnamefont {T.}~\bibnamefont {Kubo}}, \bibinfo {author} {\bibfnamefont {G.~S.}\ \bibnamefont {Lee}}, \bibinfo {author} {\bibfnamefont {H.~S.}\ \bibnamefont
  {Lee}}, \bibinfo {author} {\bibfnamefont {M.}~\bibnamefont {Matsushita}}, \bibinfo {author} {\bibfnamefont {T.}~\bibnamefont {Motobayashi}}, \bibinfo {author} {\bibfnamefont {T.}~\bibnamefont {Ohnishi}}, \bibinfo {author} {\bibfnamefont {N.~A.}\ \bibnamefont {Orr}}, \bibinfo {author} {\bibfnamefont {H.}~\bibnamefont {Otsu}}, \bibinfo {author} {\bibfnamefont {T.}~\bibnamefont {Sako}}, \bibinfo {author} {\bibfnamefont {H.}~\bibnamefont {Sakurai}}, \bibinfo {author} {\bibfnamefont {Y.}~\bibnamefont {Satou}}, \bibinfo {author} {\bibfnamefont {T.}~\bibnamefont {Sumikama}}, \bibinfo {author} {\bibfnamefont {H.}~\bibnamefont {Takeda}}, \bibinfo {author} {\bibfnamefont {S.}~\bibnamefont {Takeuchi}}, \bibinfo {author} {\bibfnamefont {R.}~\bibnamefont {Tanaka}}, \bibinfo {author} {\bibfnamefont {Y.}~\bibnamefont {Togano}}, \ and\ \bibinfo {author} {\bibfnamefont {K.}~\bibnamefont {Yoneda}},\ }\href {\doibase 10.1103/PhysRevC.93.014613} {\bibfield  {journal} {\bibinfo  {journal} {Phys. Rev. C}\ }\textbf {\bibinfo
  {volume} {93}},\ \bibinfo {pages} {014613} (\bibinfo {year} {2016})}\BibitemShut {NoStop}%
\bibitem [{\citenamefont {Fossez}\ \emph {et~al.}(2016)\citenamefont {Fossez}, \citenamefont {Rotureau}, \citenamefont {Michel}, \citenamefont {Liu},\ and\ \citenamefont {Nazarewicz}}]{PhysRevC.94.054302}%
  \BibitemOpen
  \bibfield  {author} {\bibinfo {author} {\bibfnamefont {K.}~\bibnamefont {Fossez}}, \bibinfo {author} {\bibfnamefont {J.}~\bibnamefont {Rotureau}}, \bibinfo {author} {\bibfnamefont {N.}~\bibnamefont {Michel}}, \bibinfo {author} {\bibfnamefont {Q.}~\bibnamefont {Liu}}, \ and\ \bibinfo {author} {\bibfnamefont {W.}~\bibnamefont {Nazarewicz}},\ }\href {\doibase 10.1103/PhysRevC.94.054302} {\bibfield  {journal} {\bibinfo  {journal} {Phys. Rev. C}\ }\textbf {\bibinfo {volume} {94}},\ \bibinfo {pages} {054302} (\bibinfo {year} {2016})}\BibitemShut {NoStop}%
\bibitem [{\citenamefont {Baumann}\ \emph {et~al.}(2007)\citenamefont {Baumann}, \citenamefont {Amthor}, \citenamefont {Bazin}, \citenamefont {Brown}, \citenamefont {III}, \citenamefont {Gade}, \citenamefont {Ginter}, \citenamefont {Hausmann}, \citenamefont {Mato{\v{s}}}, \citenamefont {Morrissey}, \citenamefont {Portillo}, \citenamefont {Schiller}, \citenamefont {Sherrill}, \citenamefont {Stolz}, \citenamefont {Tarasov},\ and\ \citenamefont {Thoennessen}}]{Baumann2007}%
  \BibitemOpen
  \bibfield  {author} {\bibinfo {author} {\bibfnamefont {T.}~\bibnamefont {Baumann}}, \bibinfo {author} {\bibfnamefont {A.~M.}\ \bibnamefont {Amthor}}, \bibinfo {author} {\bibfnamefont {D.}~\bibnamefont {Bazin}}, \bibinfo {author} {\bibfnamefont {B.~A.}\ \bibnamefont {Brown}}, \bibinfo {author} {\bibfnamefont {C.~M.~F.}\ \bibnamefont {III}}, \bibinfo {author} {\bibfnamefont {A.}~\bibnamefont {Gade}}, \bibinfo {author} {\bibfnamefont {T.~N.}\ \bibnamefont {Ginter}}, \bibinfo {author} {\bibfnamefont {M.}~\bibnamefont {Hausmann}}, \bibinfo {author} {\bibfnamefont {M.}~\bibnamefont {Mato{\v{s}}}}, \bibinfo {author} {\bibfnamefont {D.~J.}\ \bibnamefont {Morrissey}}, \bibinfo {author} {\bibfnamefont {M.}~\bibnamefont {Portillo}}, \bibinfo {author} {\bibfnamefont {A.}~\bibnamefont {Schiller}}, \bibinfo {author} {\bibfnamefont {B.~M.}\ \bibnamefont {Sherrill}}, \bibinfo {author} {\bibfnamefont {A.}~\bibnamefont {Stolz}}, \bibinfo {author} {\bibfnamefont {O.~B.}\ \bibnamefont {Tarasov}}, \ and\ \bibinfo {author}
  {\bibfnamefont {M.}~\bibnamefont {Thoennessen}},\ }\href {\doibase 10.1038/nature06213} {\bibfield  {journal} {\bibinfo  {journal} {Nature}\ }\textbf {\bibinfo {volume} {449}},\ \bibinfo {pages} {1022} (\bibinfo {year} {2007})}\BibitemShut {NoStop}%
\bibitem [{\citenamefont {Xie}\ \emph {et~al.}(2023)\citenamefont {Xie}, \citenamefont {Li}, \citenamefont {Michel}, \citenamefont {Li}, \citenamefont {Wang}, \citenamefont {Ong},\ and\ \citenamefont {Zuo}}]{XIE2023137800}%
  \BibitemOpen
  \bibfield  {author} {\bibinfo {author} {\bibfnamefont {M.}~\bibnamefont {Xie}}, \bibinfo {author} {\bibfnamefont {J.}~\bibnamefont {Li}}, \bibinfo {author} {\bibfnamefont {N.}~\bibnamefont {Michel}}, \bibinfo {author} {\bibfnamefont {H.}~\bibnamefont {Li}}, \bibinfo {author} {\bibfnamefont {S.}~\bibnamefont {Wang}}, \bibinfo {author} {\bibfnamefont {H.}~\bibnamefont {Ong}}, \ and\ \bibinfo {author} {\bibfnamefont {W.}~\bibnamefont {Zuo}},\ }\href {\doibase https://doi.org/10.1016/j.physletb.2023.137800} {\bibfield  {journal} {\bibinfo  {journal} {Physics Letters B}\ }\textbf {\bibinfo {volume} {839}},\ \bibinfo {pages} {137800} (\bibinfo {year} {2023})}\BibitemShut {NoStop}%
\bibitem [{\citenamefont {Mengran}\ \emph {et~al.}(2024)\citenamefont {Mengran}, \citenamefont {Jianguo}, \citenamefont {NICOLAS}, \citenamefont {Honghui},\ and\ \citenamefont {Wei}}]{publisher/Science}%
  \BibitemOpen
  \bibfield  {author} {\bibinfo {author} {\bibfnamefont {X.}~\bibnamefont {Mengran}}, \bibinfo {author} {\bibfnamefont {L.}~\bibnamefont {Jianguo}}, \bibinfo {author} {\bibfnamefont {M.}~\bibnamefont {NICOLAS}}, \bibinfo {author} {\bibfnamefont {L.}~\bibnamefont {Honghui}}, \ and\ \bibinfo {author} {\bibfnamefont {Z.}~\bibnamefont {Wei}},\ }\href {http://www.sciengine.com/publisher/Science China Press/journal/SCIENCE CHINA Physics, Mechanics & Astronomy///10.1007/s11433-023-2227-5} {\bibfield  {journal} {\bibinfo  {journal} {SCIENCE CHINA Physics, Mechanics \& Astronomy}\ }\textbf {\bibinfo {volume} {67}},\ \bibinfo {pages} {212011} (\bibinfo {year} {2024})}\BibitemShut {NoStop}%
\bibitem [{\citenamefont {Tostevin}\ and\ \citenamefont {Brown}(2006)}]{PhysRevC.74.064604}%
  \BibitemOpen
  \bibfield  {author} {\bibinfo {author} {\bibfnamefont {J.~A.}\ \bibnamefont {Tostevin}}\ and\ \bibinfo {author} {\bibfnamefont {B.~A.}\ \bibnamefont {Brown}},\ }\href {\doibase 10.1103/PhysRevC.74.064604} {\bibfield  {journal} {\bibinfo  {journal} {Phys. Rev. C}\ }\textbf {\bibinfo {volume} {74}},\ \bibinfo {pages} {064604} (\bibinfo {year} {2006})}\BibitemShut {NoStop}%
\bibitem [{\citenamefont {Nakada}\ and\ \citenamefont {Takayama}(2018)}]{PhysRevC.98.011301}%
  \BibitemOpen
  \bibfield  {author} {\bibinfo {author} {\bibfnamefont {H.}~\bibnamefont {Nakada}}\ and\ \bibinfo {author} {\bibfnamefont {K.}~\bibnamefont {Takayama}},\ }\href {\doibase 10.1103/PhysRevC.98.011301} {\bibfield  {journal} {\bibinfo  {journal} {Phys. Rev. C}\ }\textbf {\bibinfo {volume} {98}},\ \bibinfo {pages} {011301(R)} (\bibinfo {year} {2018})}\BibitemShut {NoStop}%
\end{thebibliography}%





\end{document}